\documentclass[aip,jcp,floatfix,reprint]{revtex4-1}

\usepackage[T1]{fontenc}       

\usepackage{ae,aecompl}
\usepackage{graphicx}
\usepackage{epstopdf}
\usepackage{color}
\usepackage{amsmath}
\usepackage{hyperref}

\newcommand{\comment}[1]{}

\global\long\def\M#1{\boldsymbol{#1}}

\global\long\def\sinc{\text{sinc}}

\begin{document}

\title{Rapid Sampling of Stochastic Displacements in Brownian Dynamics Simulations}

\author{Andrew M. Fiore}
\affiliation{Department of Chemical Engineering, Massachusetts Institute of Technology, Cambridge, MA 02139, USA}
\author{Florencio Balboa Usabiaga}
\author{Aleksandar Donev}
\affiliation{Courant Institute of Mathematical Sciences, New York University, New York, NY 10012}
\author{James W. Swan}
\email{jswan@mit.edu}
\affiliation{Department of Chemical Engineering, Massachusetts Institute of Technology, Cambridge, MA 02139, USA}


\begin{abstract}

We present a new method for sampling stochastic displacements in Brownian Dynamics (BD) simulations of colloidal scale particles. The method relies on a new formulation for Ewald summation of the Rotne-Prager-Yamakawa (RPY) tensor, which guarantees that the real-space and wave-space contributions to the tensor are independently symmetric and positive-definite for all possible particle configurations. Brownian displacements are drawn from a superposition of two independent samples: a wave-space (far-field or long-ranged) contribution, computed using techniques from fluctuating hydrodynamics and non-uniform Fast Fourier Transforms; and a real-space (near-field or short-ranged) correction, computed using a Krylov subspace method. 
The combined computational complexity of drawing these two independent samples scales linearly with the number of particles. The proposed method circumvents the super-linear scaling exhibited by all known iterative sampling methods applied directly to the RPY tensor that results from the power law growth of the condition number of tensor with the number of particles. For geometrically dense microstructures (fractal dimension equal three), the performance is independent of volume fraction, while for tenuous microstructures (fractal dimension less than three), such as gels and polymer solutions, the performance improves with decreasing volume fraction. This is in stark contrast with other related linear-scaling methods such as the Force Coupling Method and the Fluctuating Immersed Boundary method, for which performance degrades with decreasing volume fraction.  Calculations for hard sphere dispersions and colloidal gels are illustrated and used to explore the role of microstructure on performance of the algorithm. In practice, the logarithmic part of the predicted scaling is not observed and the algorithm scales linearly for up to $ 4 \times 10^6 $ particles, obtaining speed ups of over an order of magnitude over existing iterative methods, and making the cost of computing Brownian displacements comparable the cost of computing deterministic displacements in BD simulations. A high-performance implementation employing non-uniform fast Fourier transforms implemented on graphics processing units and integrated with the software package HOOMD-blue is used for benchmarking.
\end{abstract}

\maketitle

\section{Introduction}

	Complex fluids composed of colloidal particles or dissolved polymers (herein referred to simply as particles) are common to many industrial applications. The dynamics of these materials are determined by the structure and interactions of the constituent elements. Micron and sub-micron sized constituents interact generically through the solvent via hydrodynamic interactions (HI). HI are critical to describing large scale collective motions abundant in diffusing\cite{swan2011}, aggregating\cite{zsigi,zsigi2} and sedimenting dispersions \cite{ladd_sedimentation}. For the slow viscous flows characteristic of colloidal motion, the HIs are determined by the viscous response of the fluid to motion of the immersed particles, as described by the Stokes equations.  To account for Brownian motion of suspended particles, a fluctuating stress acting on the fluid must be incorporated into its momentum balance. On time scales that are long relative to the momentum relaxation time of the fluid, this stochastic stress has an equivalent representation as instantaneously correlated stochastic forces acting on directly on the particles.  This suggests two different approaches for modeling stochastic motion of colloids.  In one approach, the motion of the solvent is represented implicitly.  The particle motion induced by hydrodynamic forces is constructed from multipolar expansions of the fundamental solutions of the Stokes equations, and stochastic forces acting directly on the particles are used to determine the particle trajectories.  In an alternative approach, the motion of the solvent is determined explicitly either via discretization and solution of the Stokes equations around the particles, or by representing the solvent as a discrete collection of momentum carrying quasi-particles.  For these explicit solvent techniques, the stochastic stress in the fluid domain or stochastic forces acting on the suspended particles and quasi-particles represent thermal fluctuations that satisfy fluctuation dissipation balance.
	     
	The most common implicit solvent method is Brownian Dynamics (BD) with hydrodynamic interactions (BD-HI), which approximates fluid-mediated particle interactions with the Rotne-Prager-Yamakawa (RPY) tensor \cite{rpy} and explicitly integrates out inertia by taking the overdamped limit. Hydrodynamic interactions (HI) between particles described by the RPY tensor represent particles as a point force plus a quadrupole. Higher multipole moments (e.g. torques and stresslets) can be accounted for with a systematic multipole expansion, as done in Stokesian Dynamics\cite{sd}. Particle motion is linearly coupled to these multipole moments by the mobility tensor, $\mathcal{M}$. Fluctuation dissipation requires the Brownian displacements to have zero mean and variance proportional to $\mathcal{M}$, necessitating the computation of the action of $\mathcal{M}^{1/2}$ on a vector. 
Modern methods can compute the product of $\mathcal{M}$ with a vector in $\mathcal{O}\left( N \right)$ time, where $ N $ is the number of particles being modeled.\cite{RPY_FMM} The long-standing approach to multiplying  $\mathcal{M}^{1/2}$ by a vector has been to use either Cholesky decomposition or the iterative approximation of \citeauthor{fixman}\cite{fixman}. An iterative scheme with better error control has recently been developed \cite{chow-saad}, but all known iterative schemes have superlinear scaling of computational complexity with number of particles \cite{ando-krylovHI,tea} because the condition number of $\mathcal{M}$ increases as a power law in $N$. This leads to superlinear scaling for stochastic simulations using implicit solvent techniques of $\mathcal{O}\left(N^{1.25-1.50} \log N\right)$, restricting dynamic simulations to systems of no more than a few thousand particles. A technique with linear scaling is clearly needed to perform multiscale simulations of structured materials. We describe here a novel linear-scaling algorithm for calculating Brownian motion in hydrodynamically interacting systems.

      	Explicit solvent methods which include fluid inertia, the most common of which are Lattice Boltzmann (LB) \cite{lb1,lb2,lb3}, Stochastic Rotation Dynamics (SRD) \cite{mpc}, and Dissipative Particle Dynamics (DPD) \cite{dpd1,dpd2}, are commonly cited as having linear scaling in the number of particles, even for stochastic simulations. While the operation count of explicit solvent algorithms is linear in the number of particles, these algorithms are limited by the need to resolve inertial degrees of freedom in the solvent, which results in a significant loss of efficiency on approach to overdamped systems. To approximately maintain overdamped dynamics at all simulated length scales, the dimensionless numbers relating process time scales to the rate of vorticity diffusion over the system size must also be controlled. For example, in SRD careful analysis has been performed to determine the relationship between simulation parameters and dimensionless groups such as the Schmidt, Reynolds, and Mach Numbers \cite{srd_scaling}. Controlling these variables over all length scales is especially important for flow and self-assembly in structured systems, where particles can form aggregates or percolating networks that approach the system size. The Reynolds number can be written as $Re=\ell^{2}/\tau\nu$ where $\ell$ is the relevant length scale, $\nu$ is the kinematic viscosity of the fluid, and $\tau$ is the flow time scale. An overdamped simulation of colloids, nanoparticles, or polymers should be performed so that $L^{2}/\tau\nu < Re_{c}$ where $Re_{c}$ is the maximum allowable Reynolds number and $L$ is the system size. The simulation time step $\Delta t$ must be proportional to $\tau_{I}=a^2/\nu$ to resolve vorticity diffusion, where $a \sim \left(L^{3}\phi/N\right)^{1/3}$ is the particle size and $\phi$ is the volume fraction of particles in the system. Therefore, the number of time steps needed to resolve the flow time scale is 
$\tau / \tau_{I} \sim \phi^{-2/3}\,Re_{c}^{-1}N^{2/3}$.
Each time step requires $\mathcal{O}\left(N\right)$ operations, and thus the overall computational complexity of explicit solvent techniques with control over vorticity diffusion at all length scales is $\mathcal{O}\left(N^{5/3}\right)$. Another way of thinking about this scaling is that $\mathcal{O}\left(V\right) \sim N$ operations are required per time step to solve the Stokes equations, but $L^{2}/\nu \sim N^{2/3}$ time steps required for vorticity to diffuse throughout the system volume, leading to the $\mathcal{O}\left(N^{5/3}\right)$ scaling.

A class of ``explicit solvent'' methods directly tackles the overdamped limit by solving the {\em steady} Stokes equations\cite{fib,fcm,ForceCoupling_Lubrication,FluctuatingFCM_DC}. This approach avoids using an analytic form of the mobility tensor, which can be an advantage for non-periodic domains\cite{fib}. Furthermore, the Brownian displacements needed for stochastic simulations can be generated with little to no extra cost, as demonstrated with the development of the Fluctuating Immersed Boundary (FIB) method,\cite{fib} and later incorporated into the fluctuating Force Coupling Method (FCM)\cite{fcm,FluctuatingFCM_DC} (which also includes higher-order stresslet corrections in the computation of the hydrodynamic interactions).
This is accomplished by using fluctuating hydrodynamics, which in this context amounts to adding a stochastic forcing in the Stokes equations in the form of a stochastic stress tensor, rather than adding ``Brownian forces'' to the particles as in more traditional approaches. Our Positively-Split-Ewald (PSE) method builds on this key idea to compute the far-field (long-ranged) contribution to the hydrodynamic interactions in a periodic domain, by using Fast Fourier Transforms (FFTs) to efficiently solve the fluctuating Stokes equation.\cite{fib,fcm} 

Both the FCM and FIB methods, while $\mathcal{O}\left(N\right)$, share a number of important shortcomings. The first of these is that in FCM/FIB one cannot independently control the form of the {\em near-field} hydrodynamic interactions between particles. Specifically, an RPY-like hydrodynamic mobility is obtained at large distances, but at short distances the hydrodynamic interaction tensor is different from RPY, in a way that depends on the specific form of the kernel (envelope) function used to transfer forces from the particles to the fluid. While one can argue that the RPY tensor is only accurate in the far field anyway, it is troublesome from a numerical methods perspective to change the physics of the model in order to make the numerical computations easier. The second important shortcoming is that methods such as FCM/FIB require a grid whose spacing is on the order of the particle size or smaller. This means that at low densities, where the RPY approximation is most accurate, these methods require large grids that eventually cannot even fit in the computer memory. FIB is better than FCM in this respect, since it allows for very compact kernels (3 or 4 grid points only in each direction) through the use of Peskin's specially-designed immersed boundary kernels.\cite{peskin-ib}
However, a notable shortcoming of FIB, and in fact of all particle-mesh Ewald approaches as well, is that the accuracy of the computations cannot be controlled independently and some degree of lack of translational invariance must be accepted and then corrected for.\cite{fib} 

Our PSE method overcomes the super-linear scaling of BD-like methods while bypassing the shortcomings of FCM/FIB-like methods and generates Brownian displacements whose variance is proportional to the exact, periodic RPY tensor. The PSE method represents the RPY tensor as a sum of two symmetric positive definite operators. We call this ``positive splitting" because both parts of the RPY tensor are independendently positive-definite for all particle configurations. One operator is spatially local and easily evaluated via an exact analytical formula as in BD-like methods.  The other operator is non-local and easily evaluated via FCM/FIB-like methods. The accuracy of the non-local operator is controlled by building on the recently-developed Spectral Ewald (SE) method,\cite{lindbo-tornberg} which decouples numerical errors in interpolation from numerical errors resulting from discretization and solution of the steady Stokes equations.  Samples of the Brownian displacement are drawn from a sum of two independent displacements whose variances are proportional to the local and the global parts of the RPY tensor, respectively.  Local contributions are computed using the recently-proposed iterative Lanczos method for evaluating the action of the square root of a matrix.\cite{chow-saad}  Non-local contributions are computing using an approach inspired by fluctuating hydrodynamics. We show that this method is $\mathcal{O}(N)$ and that the accuracy of the various numerical approximations is completely decoupled. We can therefore optimize the algorithm to maximize efficiency, {\em without} sacrificing accuracy, which is controlled independently by a chosen relative error tolerance. In this sense the discretization, interpolation, and iterative approximations used in the PSE method are merely computational tools that do not affect the physics (RPY mobility) to within a prescribed tolerance.

The manuscript is organized as follows.  In section IIA, we develop the PSE method by deriving a new Fourier space representation of the RPY tensor in a periodic geometry.  Then, we utilize Ewald summation to deconstruct this tensor into local and non-local parts.  In section IIB, we demonstrate how samples of the Brownian displacement may be drawn independently from two normal distributions with variance proportional to the local and non-local parts of the RPY tensor.  In section IIC, we discuss optimization of algorithm performance while controlling the numerical error in all parts of the proposed algorithm.  In Section III, we present various numerical tests which characterize the overall algorithm performance and compare that performance to other standard algorithms as well as analytically derived performance estimates.

\section{Positively Split Ewald approach to Brownian Dynamics}

	\subsection{The Mobility Tensor}
        
	The primary difficulty in performing dynamic simulations of hydrodynamically interacting particles is to calculate the velocity imparted to each particle due to forces acting on the other particles. In colloidal systems, the dynamics are overdamped, so the fluid motion is described by the Stokes equations. The fluid-mediated coupling among particles is represented by the symmetric, positive-definite mobility tensor $\mathcal{M}$. In a very common approach, $\mathcal{M}$ is constructed using a multipole expansion of the force density on particle surfaces to determine the flow field produced by the particles. Particle motion is inferred from Fax\'{e}n laws that dictate how a flow field exerts stresses on the particle surfaces. Evaluating the multipole expansion requires knowledge of the Green's function for Stokes flow subject to known boundary conditions. 
        
        Periodic systems are common in simulation, and the appropriate Green's function is given by \citeauthor{hasimoto}\cite{hasimoto}
	\begin{equation}
		{\bf J}\left({\bf x},{\bf y}\right) = \frac{1}{\eta V}\sum_{{\bf k}\neq 0}e^{i{\bf k}\cdot({\bf x}-{\bf y})}\frac{1}{k^{2}}\left({\bf I}-\hat{\bf k}\hat{\bf k}\right), \label{eqn:green}
	\end{equation}
	where ${\bf J}\left({\bf x},{\bf y}\right)\cdot{\bf F}$ is the fluid velocity at point ${\bf x}$ propagated by a force ${\bf F}$ located at point ${\bf y}$ and its periodic images on a lattice, ${\bf I}$ is the identity tensor, ${\bf k}$ are drawn from the set of reciprocal space lattice vectors excluding the zero vector, $\eta$ is the fluid viscosity, and $V$ is the periodic cell volume. The sum over periodic images in Equation \eqref{eqn:green} is slowly converging in most cases and divergent for $ \bf x = \bf y $. For use in practical computation the summation process can be accelerated by using an Ewald method.  A splitting function, $H(k,\xi)=\left( 1 + \frac{k^{2}}{4\xi^{2}}\right)e^{-k^{2}/4\xi^{2}}$, introduced by \citeauthor{hasimoto}\cite{hasimoto} is used to decompose the sum into real space and wave space parts
	\begin{align}
		{\bf J}\left({\bf x},{\bf y}\right) &= \frac{1}{4\pi\eta}\sum\limits_{\bf R}\left[ \frac{\xi}{\pi^{1/2}} \, \phi_{-1/2}\left(\xi^{2}r^{2}\right)\,{\bf I} \right. \nonumber \\
		&\left. - \frac{\xi^{3}r^{2}}{\pi^{1/2}} \, \phi_{1/2}\left(\xi^{2}r^{2} \right) \, \left( {\bf I} - \hat{\bf r}\hat{\bf r} \right) \right]  \nonumber \\
										   &+ \frac{1}{\eta V}\sum_{{\bf k}\neq 0}e^{i{\bf k}\cdot({\bf x}-{\bf y})}\frac{1}{k^{2}} H\left(k,\xi\right)\left({\bf I}-\hat{\bf k}\hat{\bf k}\right), \label{eqn:green2}
	\end{align}		
	where ${\bf R}$ is the set of real space lattice vectors, ${\bf r} = {\bf x}-{\bf y}+{\bf R}$, and $\phi_{\nu}$ are the incomplete gamma functions. The error $\epsilon$ associated with truncating the sums at finite radii $r_{\rm cut}$ and $k_{\rm cut}$ decays exponentially with a rate controlled by the splitting parameter, $\xi$. The real space and wave space truncation errors decay as $\epsilon_{R}\sim e^{-\xi^{2}r_{\rm cut}^{2}}$ and $\epsilon_{W}\sim e^{-k_{\rm cut}^{2}/4\xi^{2}}$, respectively.
        
        Assuming rigid spherical particles of equal radius $a$, a multipole expansion using this Green's function along with Fax\'en's first formula yields a periodic form of the well-known Rotne-Prager (RP) tensor:\cite{rpy,SD_SpectralEwald}
	\begin{equation}\label{RP_differential}
		{\bf M}^{ij}_{\rm RP} = \left\{ \begin{array}{ll} \frac{1}{6\pi \eta a}, & i = j \\ \left[ \left( 1 + \frac{a^{2}}{6}\nabla_{{\bf x}}^{2} \right)\left( 1 + \frac{a^{2}}{6}\nabla_{{\bf y}}^{2} \right) {\bf J}\left({\bf x},{\bf y}\right) \right]_{\substack{{\bf x}={\bf x}_{i},\\{\bf y}={\bf x}_{j}}}, & i \ne j \end{array} \right.
	\end{equation}
	where ${\bf x}_{i}$ is the location of the center of particle $i$, and ${\bf U}_{i} = {\bf M}^{ij}_{\rm RP} \cdot {\bf F}_{j}$ is the velocity of particle $i$ induced by the force ${\bf F}_{j}$ on particle $j$. However, it is known that the operator \eqref{RP_differential} is symmetric positive definite (SPD) only when particles are not closer than the sum of their hydrodynamic radii. Additional regularizing corrections to maintain positivity are needed for overlapping configurations, which can occur in e.g. simulations of coarse-grained polymer chains or polymer-coated colloids. The common form used in BD-HI simulations is the Rotne-Prager-Yamakawa (RPY) tensor, which is SPD for all particle configurations, and has the form of an isotropic tensor for an unbounded domain:\cite{rpy}
\begin{equation} 
\label{RPY_free_space}
{\bf M}^{ij}_{\rm RPY}=\frac{1}{6\pi\eta a}\begin{cases}
\left( \frac{3a}{4r}+\frac{a^{3}}{2r^{3}} \right)\M I+ \left( \frac{3a}{4r}-\frac{3a^{3}}{2r^{3}}\right) \hat{\bf r}\hat{\bf r}, &  r>2a\\
\left( 1-\frac{9r}{32a} \right)\M I+ \left( \frac{3r}{32a} \right) \hat{\bf r}\hat{\bf r}, &  r\leq2a
\end{cases},       
\end{equation}
where ${\bf r} = {\bf x}_{i}-{\bf x}_{j}$. For periodic domains, a compact form of the RPY tensor must be derived.

Starting with the periodic Green's function, it is possible to derive a relationship for the mobility that is positive definite for all particle separations with no special regularization required, and whose constituent parts (the real space and wave space sums) are always independently positive-definite. This approach is based on the integral formulations of the the Stokes equations and integral representations of the Fax\'{e}n laws for rigid bodies:
	\begin{equation}
		{\bf u}({\bf x}) = \int_{S} dS({\bf y}) \, {\bf J}({\bf x},{\bf y}) \, \cdot {\bf f}({\bf y}), \label{eqn:ib}
	\end{equation}
	\begin{equation}
		{\bf U} = \frac{1}{4\pi a^{2}}\int_{S} dS({\bf x}) \, {\bf u}^{\prime}({\bf x}), \label{eqn:faxen}
	\end{equation}
	where ${\bf u}$ is the velocity field produced by a force density ${\bf f}$ on the particle surface $S$, and ${\bf U}$ is the velocity at which a particle moves in response to a flow field ${\bf u}^{\prime}$ on its surface, induced by the other particles. Combining equations \eqref{eqn:ib} and \eqref{eqn:faxen} for spheres of equal sizes, and keeping only the first moment in the multipole expansion of the force density, ${\bf f} = {\bf F}/4\pi a^{2}$, gives a velocity-force mobility relation between two particles,
	\begin{equation}
		{\bf U}_{i} = \frac{1}{4\pi a^{2}}\int_{S_{i}} dS_{i}({\bf x}) \, \frac{1}{4\pi a^{2}}\int_{S_{j}} dS_{j}({\bf y}) \, {\bf J}({\bf x},{\bf y}) \, \cdot {\bf F}_j,
	\end{equation}
	where ${\bf x}$ is on the surface of particle $i$ and ${\bf y}$ is on the surface of particle $j$. As discussed in Refs.~\onlinecite{RPY_Shear_Wall,RPY_Periodic_Shear}, this representation is identical to \eqref{RP_differential} for non-overlapping particles, but also gives an SPD tensor for overlapping particles as well, and can be used to generalize the RPY tensor to other geometries. Using \eqref{eqn:green} and evaluating the integrals gives the pair mobility for spherical particles with equal finite size,
	\begin{equation}
		{\bf M}_{ij} = \frac{1}{\eta V}\sum_{{\bf k}\neq 0}e^{i{\bf k}\cdot({\bf x}_{i}-{\bf x}_{j})}\frac{1}{k^{2}}\left(\frac{\sin ka}{ka}\right)^{2}\left({\bf I}-\hat{\bf k}\hat{\bf k}\right). \label{eqn:pairmobility}
	\end{equation}
	This mobility describes the hydrodynamic interaction between two finite spheres and is positive definite and unconditionally convergent for all particle separations. The $\sinc (ka) = (\sin ka) / ka$ terms are shape factors that account for the way in which spheres propagate flows into the fluid. Performing the calculations in \eqref{RP_differential} on the Hasimoto Green's function \eqref{eqn:green}, produces a shape factor that is instead $\left(1-k^{2}a^{2}/6\right)^{2}$, which matches the Taylor expansion of $\sinc(ka)$ up to second order in $k$. Higher order terms in the expansion of $\sinc (ka)$ vanish after integration over the wave vectors due to the biharmonic nature of the Stokes flow. Equation \eqref{eqn:pairmobility} appears to be a very simple yet previously unknown Fourier space representation of the RPY tensor that will be the basis of our positive Ewald splitting; if one replaces the sum by an integral over all ${\bf k}$, \eqref{eqn:pairmobility} gives the standard free-space RPY tensor \eqref{RPY_free_space}.

\subsubsection{Positively Split Ewald RPY tensor}

        The first component required to perform BD-HI efficiently is to have a fast method for multiplying the mobility matrix by a vector. For the RPY tensor in an unbounded domain, a Fast Multipole Method (FMM) has been developed by rewriting the real-space representation of the far-field RPY tensor as a combination of several scalar (charge) monopole and dipole FMM evaluations.\cite{RPY_FMM} In this approach the near-field RPY correction is simply incorporated in the near-field sum in the FMM. For point Stokeslets (monopoles), which do not include the quadrupolar far-field correction of the RPY tensor, an alternative fast Spectral Ewald method based on using the FFT has been developed by Lindbo and Tornberg\cite{lindbo-tornberg} by building on the non-uniform FFT (NUFFT) method.\cite{NUFFT} This method was recently incorporated into a Stokesian Dynamics (SD) method for Brownian polydisperse sphere suspensions by \citeauthor{SD_SpectralEwald}\cite{SD_SpectralEwald}. The SD mobility includes the RP tensor, but the approach proposed by \citeauthor{SD_SpectralEwald}\cite{SD_SpectralEwald} applies the differential formulation of the Fax\'{e}n laws, which yields a non-positive splitting even for well-separated particles. Therefore, inclusion of Brownian motion is not straightforward. Here we propose a positive Ewald splitting of the RPY tensor that allows us to use the Spectral Ewald approach to efficiently and accurately evaluate the far-field contribution and rapidly generate Brownian increments.

	Applying the Ewald sum splitting of \citeauthor{hasimoto}\cite{hasimoto} to decompose Equation \eqref{eqn:pairmobility} as ${\bf M}_{ij} = {\bf M}_{ij}^{(w)} + {\bf M}_{ij}^{(r)}$ into a long-ranged wave-space part ${\bf M}_{ij}^{(w)}$, and a short-ranged real-space part ${\bf M}_{ij}^{(r)}$, and analytically performing the integral for the inverse Fourier transform to the short-ranged part, we get
	\begin{equation}
		{\bf M}_{ij}^{(w)} = \frac{1}{\eta V}\sum_{{\bf k}\neq 0}e^{i{\bf k}\cdot({\bf x}-{\bf y})}\frac{\sinc^2\left(ka\right)}{k^{2}}  H(k,\xi) \left({\bf I}-\hat{\bf k}\hat{\bf k}\right), \label{eqn:Mwave}
	\end{equation}
	\begin{equation}
		{\bf M}_{ij}^{(r)} = F(r,\xi) \, \left( {\bf I} - \hat{\bf r}\hat{\bf r} \right) + G(r,\xi) \, \hat{\bf r}\hat{\bf r}, \label{eqn:Mreal}
	\end{equation}        
        where
        \begin{equation}
        H(k,\xi) = \left(1+\frac{k^{2}}{4\xi^{2}}\right)e^{-k^{2}/4\xi^{2}},
        \end{equation}
	and $F(r,\xi)$ and $G(r,\xi)$ are scalar functions given by
	\begin{align}
	F(r,\xi) &= \frac{1}{\eta} \frac{1}{2\pi^{2}}\int\limits_{0}^{\infty} dk \, \left( 1 - H\left( k,\xi \right) \right) \sinc^{2}\left(k a\right) \nonumber \\
	& \times \left( \frac{kr \cos kr + \left(k^{2}r^{2}-1\right)\sin kr}{k^{3}r^{3}} \right), \label{eqn:int1}
	\end{align}
	\begin{align}
	G(r,\xi) &= \frac{1}{\eta} \frac{1}{\pi^{2}}\int\limits_{0}^{\infty} dk \, \left( 1 - H\left( k,\xi \right) \right) \sinc^{2}\left(k a\right) \nonumber\\
	& \times \left( \frac{ \sin kr - kr \cos kr}{k^{3}r^{3}} \right). \label{eqn:int2}
	\end{align}
	The integrals in equations \eqref{eqn:int1} and \eqref{eqn:int2} have complicated closed-form solutions, which are given in Appendix \ref{app:RealSpace}. 
        
	These pairwise mobility functions can be easily applied to many-body simulations by summing the pairwise interactions:
	\begin{equation}
		{\bf U}_{i} = \sum\limits_{j} \left({\bf M}_{ij}^{(w)} + {\bf M}_{ij}^{(r)}\right)\cdot {\bf F}_{j}.
	\end{equation}
	This set of sums can be written as the grand mobility tensor, $\mathcal{M}$
	\begin{equation}
		{\bf U} = \mathcal{M} \cdot {\bf F},
	\end{equation}
	where ${\bf U}$ and ${\bf F}$ are vectors containing the velocity and force on all particles, respectively, and the blocks of $\mathcal{M}$ correspond to ${\bf M}_{ij}$. Like the pairwise mobility, the grand mobility has real space and wave space components, $\mathcal{M} = \mathcal{M}^{(w)} + \mathcal{M}^{(r)}$. $\mathcal{M}$ is symmetric and positive definite for all particle configurations. Using our approach, $\mathcal{M}^{(w)}$ and $\mathcal{M}^{(r)}$ are also guaranteed to independently be positive definite for all particle configurations and values of the splitting parameter, as demonstrated in Appendix \ref{app:SPD}. 
	
	The shape factor $s^{2}(ka) = {\rm sinc}^{2}(ka)$ and the splitting factor $H(k,\xi)$ collectively control whether a given representation of the RPY tensor can be positively split; both factors must be non-negative to guarantee positivity, which is required for the sampling method described in the next section. This is illustrated by the Ewald sum of the RPY pair mobility tensor
	\begin{align}
		{\bf M}_{\rm RPY}^{ij} &= \sum\limits_{{\bf k}\neq{\bf 0}} s^{2}(ka) \, \left(1 - H(k,\xi)\right) \, {\bf P}_{k} \nonumber \\
		&+ \sum\limits_{{\bf k}\neq{\bf 0}} s^{2}(ka) \, H(k,\xi) \, {\bf P}_{k},
	\end{align}
	 where ${\bf P}_{k} = e^{i{\bf k}\cdot\left({\bf x}_{i}-{\bf x}_{j}\right)} {k^{-2}}\left( {\bf I} - \hat{\bf k}\hat{\bf k}\right)$, and the first sum is the wave space representation of the real space sum. Both $s^{2}(ka) \, \left(1 - H(k/\xi)\right)$ and $s^{2}(ka) \, H(k/\xi)$ must be non-negative if each sum is to be SPD. Since $s^{2}(ka)$ and $H(k/\xi)$ are chosen independently, both $s^{2}(ka) \geq 0$ and $0 \leq H(k/\xi) \leq 1$ are required to guarantee a positive splitting for all values of $k$ and $\xi$, i.e. guarantee that both the real space and wave space components of the mobility tensor are independently positive-definite. Beenakker's formulation of the Ewald sum of the RP tensor uses a differential representation of the tensor, shown in \eqref{RP_differential}, and does not include overlap corrections\cite{beenakker}. 
The wave space representation of the shape factor associated with this formulation is $s^{2}_{B}(ka) = 1 - \frac{a^{2}k^{2}}{3}$, which clearly is not positive when $a^{2}k^{2} > 3$. Therefore, RPY formulations based on this shape factor, even those that account for overlap corrections, cannot ensure a positive splitting with \textit{any} splitting factor. In contrast, the shape factor used in this work is always non-negative. The two prevalent splitting factors are due to Hasimoto \cite{hasimoto}, $H_{H}(k/\xi) = \left(1+\frac{k^{2}}{4\xi^{2}}\right)e^{-k^{2}/4\xi^{2}}$, used in this work, and Beenakker, $H_{B}(k/\xi) = \left(1+\frac{k^{2}}{4\xi^{2}} + \frac{k^{4}}{8\xi^{4}}\right)e^{-k^{2}/4\xi^{2}}$, which has been used used in prior works \cite{beenakker,RPY_Periodic_Shear}. Of these two, only the Hasimoto factor satisfies the condition that $0 \leq H(k/\xi) \leq 1$. Any choice of $H(k/\xi)$ such that $0 \leq H(k/\xi) \leq 1$ can be used in principle. The Hasimoto function is convenient because it results in exponentially decaying real-space sums. Additionally, any positive shape factor, e.g. the Gaussian kernels employed in the FCM technique \cite{fcm}, can be used. The ${\rm sinc}^{2}(ka)$ shape factor derived in this work yields unconditionally convergent sums that \emph{exactly} match the Stokes drag and RPY tensor for both overlapping and non-overlapping particles, without needing to explicitly consider overlap corrections.
			
	\subsection{Brownian Dynamics with Fast Stochastic Sampling}
        
	In Brownian dynamics simulations, particle trajectories evolve according to the overdamped Langevin equation
	\begin{equation}
		d{\bf x} = \mathcal{M}\cdot{\bf F} \, dt + \sqrt{2k_{B}T} \, \mathcal{M}^{1/2} \cdot d {\bf W}\left(t\right) + k_{B}T \left(\nabla \cdot \mathcal{M}\right) dt ,\label{eqn:langevin}
	\end{equation}
	where $d{\bf x}$ is the set of particle displacements, $k_{B}$ is Boltzmann's constant, $T$ is temperature, ${\bf F}$ are forces applied to the particles, and ${\bf W}(t)$ is a collection of independent standard Wiener processes.
        Here $\mathcal{M}^{1/2}$ denotes a ``square root'' of the mobility and can be any matrix that satisfies $\mathcal{M}^{1/2} \left(\mathcal{M}^{1/2}\right)^\dagger = \mathcal{M}$, where dagger denotes an adjoint. For translationally-invariant systems, including periodic domains, the divergence of the mobility appearing in the final stochastic drift term vanishes, $\nabla \cdot \mathcal{M}=0$, and it is not necessary to use special methods in order to capture this term;\cite{fib,FluctuatingFCM_DC} one can use the Euler-Maruyama method for temporal integration. Applying the action of the square root of the mobility is the most time-consuming calculation in a simulation. Iterative methods using e.g. Chebyshev polynomials\cite{fixman} or a Lanczos process \cite{chow-saad} require $N^{0.25}-N^{0.5}$ evaluations of the mobility to evaluate the square root. The number of iterations required has a power law dependence on the condition number of $\mathcal{M}$, which is a monotonically increasing function of $N$. The total operation count of this approach is $N^{1.25-1.5}\log N$ for accelerated algorithms such as particle-mesh Ewald \cite{pme} or Spectral Ewald \cite{lindbo-tornberg}. 
        
        We propose an alternative approach utilizing our novel formulation of the mobility to draw independent samples from the real space and wave space parts of the mobility,
	\begin{equation}
		\mathcal{M}^{1/2} \cdot d {\bf W}\left(t\right) 
                \,{\buildrel d \over =}\
                \mathcal{M}_{R}^{1/2} \cdot d {\bf W}_{1}\left(t\right) + \mathcal{M}_{W}^{1/2} \cdot d {\bf W}_{2}\left(t\right), \label{eqn:split}
	\end{equation}
	where ${\bf W}_{1}$ and ${\bf W}_{2}$ are two independent Wiener processes. The two separate samples can be efficiently approximated by leveraging the structure of $\mathcal{M}_{R}$ and $\mathcal{M}_{W}$. 

\subsubsection{Real Space Contribution}
        
\begin{figure}
	  \includegraphics{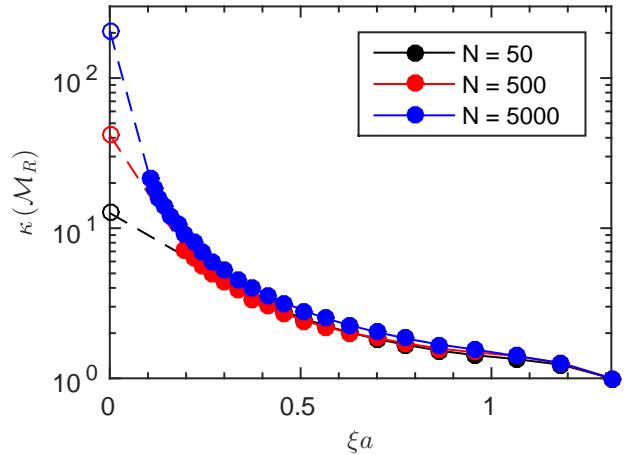}
	  \caption{Condition number of the real space mobility tensor $\mathcal{M}_{R}$ as a function of ${\xi}a$ for varying $N$.}
	  \label{fig:eigenspectrum}
	\end{figure}

        The action of $\mathcal{M}_{R}^{1/2}$ can be readily approximated by iterative schemes such as those proposed by \citeauthor{fixman}\cite{fixman} or \citeauthor{chow-saad}\cite{{chow-saad}} because of the short-ranged nature of the kernels $F$ and $G$ given in Eqs.\eqref{eqn:int1}-\eqref{eqn:int2} for {\em finite} $\xi>0$. For a sufficiently small real space cutoff radius (sufficiently large $\xi$), $\mathcal{M}_{R}$ contains only local interactions and therefore has a compact eigenspectrum. Figure \ref{fig:eigenspectrum} shows the condition number $\kappa$ of $\mathcal{M}_{R}$ as a function of $\xi$ for varying $N$. The condition number is independent of $N$ at a fixed cutoff radius, therefore the number of iterations is also asymptotically independent of $N$. The limit ${\xi}a \rightarrow 0$ represents the condition number of the full RPY mobility, $\mathcal{M}$, which contains long-ranged hydrodynamic interactions decaying as the inverse power of the distance, and becomes increasingly ill-conditioned for increasing $N$.
	
\subsubsection{Wave Space Contribution}
	
	The wave space component of the mobility can be written as a sequence of linear operations:
        \begin{equation}
        \mathcal{M}_{w} = {\bf D}^{-1} \cdot {\bf B} \cdot {\bf D},
        \end{equation}
        where the diagonal matrix ${\bf B}$ maps the wave space representation of forces to velocities, and ${\bf D}$ is the non-uniform Discrete Fourier transform (NUDFT) operator. The application of ${\bf D}$ converts the point forces applied on the particles to Fourier space. The application of ${\bf B}$ multiplies each Fourier component by the {\em non-negative} factor (see \eqref{eqn:Mwave})
        \[
        \frac{1}{\eta V}\frac{\sinc^2\left(ka\right)}{k^{2}} H(k,\xi),
        \]
        and projects onto the space of divergence free velocity fields via the projector $\left({\bf I}-\hat{\bf k}\hat{\bf k}\right)$. Lastly, the inverse NUDFT ${\bf D}^{-1}$ converts back to real space by evaluating the velocities at the positions of the particles. 
        The matrix ${\bf B}$ is block-diagonal and positive semi-definite and therefore has a trivial analytical square root. Using the unitary property of the Fourier transform allows $\mathcal{M}_{w}$ to be re-expressed as
	\begin{equation}
		\mathcal{M}_{w} = \left( {\bf D}^{\dagger} \cdot {\bf B}^{1/2} \right) \cdot \left( {\bf D}^{\dagger} \cdot {\bf B}^{1/2} \right)^{\dagger}.
	\end{equation}
        This shows that the wave space Brownian displacement can be calculated with a single matrix multiplication
	\begin{equation}
		\mathcal{M}_{w}^{1/2} \cdot d{\bf W}_{2}\left(t\right) \equiv {\bf D}^{\dagger} \cdot {\bf B}^{1/2} \cdot d{\bf W}_{2}(t).
	\end{equation}
        The random vector $d{\bf W}_{2}\left(t\right)$ is complex valued because it is a random variable in wave space, unlike $d{\bf W}_{1}\left(t\right)$ which is real valued. Some care must be exercised to ensure the proper conjugacy conditions, $d{\bf W}_{2}\left(-{\bf k}\right)=d{\bf W}_{2}^{\dagger}\left({\bf k}\right)$, and obtain a real-valued particle velocity; details of this can be found in the article describing the fluctuating FCM method.\cite{fcm}
        
        In our work, following the approach taken for point Stokeslets in the Spectral Ewald method\cite{lindbo-tornberg}, we use the non-uniform FFT (NUFFT) method of Greengard and Lee\cite{NUFFT} to obtain the action of ${\bf D}^{\dagger}$. The only change is to add the additional sinc factors in \eqref{eqn:Mwave} to give the RPY regularization of the Oseen tensor, i.e., replacing the inverse Laplacian $k^{-2}$ with $\sinc^2\left(ka\right)/k^{2}$. The Spectral Ewald approach is also described in detail and compared to alternative particle-mesh Ewald approaches in Ref. \onlinecite{SD_SpectralEwald}. Here we only give a very brief description. The basic idea is to transfer information between the particle positions and a regular grid (to be used for the FFTs) using a Gaussian kernel truncated at $m$ standard deviations, which is sufficiently resolved by the grid to obtain a desired error bound (number of digits of accuracy). The spreading of forces from the particles to the grid, and its adjoint operation, interpolation of velocities from the grid to the particles, is accomplished by directly evaluating the Gaussian spreading kernel on a subgrid of $P^3$ grid points around each particle, where $P$ is a sufficiently large integer for a given error tolerance.\cite{lindbo-tornberg} In the Spectral Ewald approach we have ${\bf D} = \mathcal{F}\cdot{\bf S}$ so $\mathcal{M}_{w} = {\bf S}^{\dagger} \cdot \mathcal{F}^{-1} \cdot {\bf B} \cdot \mathcal{F} \cdot {\bf S}$, where $\mathcal{F}$ represents the FFT. Here ${\bf S}$ spreads particle forces to the uniform grid and its adjoint ${\bf S}^{\dagger} = \sigma \, {\bf S}^{T}$ is the interpolation operator, where $\sigma$ is the volume of a grid cell. This gives the desired
        \[
        \mathcal{M}_{w}^{1/2} \equiv \sigma^{1/2} \, {\bf S}^{T} \cdot \mathcal{F}^{\dagger} \cdot {\bf B}^{1/2},
        \]
        which can be applied to a vector of complex-valued Gaussian variates efficiently by a combination of rescaling, inverse Fourier transform, and interpolation operations. While we have used the Spectral Ewald approach to calculate the Ewald sum because of its spectral accuracy, in principle any algorithm for evaluating the wave space sum, e.g. Particle-Mesh Ewald, could be used.
The above approach to the stochastic calculation is very similar to the one used in the FIB\cite{fib} and fluctuating FCM\cite{fcm} methods. Although presented here from an algebraic perspective, the method has a very natural physical interpretation based on fluctuating hydrodynamics. Specifically, it can be seen as generating long-range correlated particle velocities (stochastic displacements) by applying a fluctuating stress to the fluid and then interpolating the resulting random velocity field at the positions of the particles.\cite{DDFT_Hydro} As in FCM/FIB our method employs a smoothly varying kernel to represent the force density applied by the particles in the fluid. FCM uses Gaussian kernels, while FIB uses compactly supported kernels ($P=3,4,{\rm or \:\: }6$) developed for the immersed boundary method.\cite{peskin-ib,peskin-kernel} The present work uses the sinc Fourier shape factors convoluted with Gaussians as the representation of the force distribution. In the FIB approach, unlike both FCM and our PSE method, the steady Stokes equations are solved using multigrid algorithms instead of FFTs; this makes the method easy to generalize to non-periodic domains. For a periodic system with only monopole terms included, the Brownian increment generation in FCM is analogous to the current approach with the real space sum neglected. For non-overlapping particles, PSE is equivalent to FCM if one chooses a value of $\xi$ at which the error in truncating the real space sum at $r_{\rm cut}=2a$ is less than the desired tolerance. The constraint to operate at larger $\xi$ and the ability to balance calculations between real space and wave space means that the PSE approach is potentially superior to FCM for periodic simulations of Brownian motion. Specifically, the PSE operator splitting provides flexibility to optimize performance across a broad range of volume fractions and microstructures guided by Equations \eqref{eqn:xistar} and \eqref{eqn:tstar}.
	
	\subsection{Optimizing Performance}	
        
\begin{figure}
	  \includegraphics{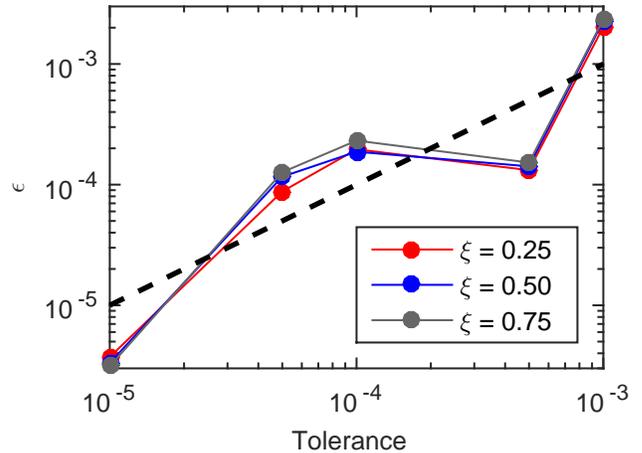}
	  \caption{Relative error in the mobility calculation as a function of desired tolerance for different $\xi$. The error is defined as $\epsilon=\lVert {\bf U}-{\bf U}_{\rm exact}\rVert / \lVert {\bf U}_{\rm exact} \rVert$, where the exact solution was computed using an error tolerance of $10^{-10}$. The dashed line indicates equality between the computed error and the specified error tolerance. Discrete steps (integer values) are used for the support size in the NUFFT quadrature operations, which causes the bump at tolerances near $10^{-4}$.}
	  \label{fig:parity}
	\end{figure}

	The algorithm's performance can be easily optimized when using the Spectral Ewald approach \cite{lindbo-tornberg} because the sources of numerical error are decoupled and carefully controlled. There are three independent contributions to the total error in the Spectral Ewald approximation of $\mathcal{M}$: truncation of the real space sum at finite radius $r_{\rm cut}$, truncation of the wave space sum at finite wave vector $k_{\rm cut}$, and quadrature. The real space and wave space truncation errors are bounded by $\epsilon_{R}\sim e^{-\xi^{2}r_{\rm cut}^{2}}$ and $\epsilon_{W}\sim e^{-k_{\rm cut}^{2}/4\xi^{2}}$. The quadrature error is bounded by $\epsilon_{q} \sim e^{-\pi^{2}P^{2}/2m^{2}} + {\rm erfc}\left(m/\sqrt{2}\right)$, where $m$ is the width of the Spectral Ewald Gaussian used in the NUFFT, and $P$ is the number of grid points in each particle's support per spatial dimension. All errors decays exponentially, so the method is spectrally accurate. Furthermore, all the errors are independent, so the bounds for each term can be used to separately choose the operating parameters $r_{\rm cut}$, $k_{\rm cut}$, $m$, and $P$. Figure \ref{fig:parity} shows the error in the sedimentation velocity for a random suspension of 5000 hard spheres ($\phi=0.2$) as a function of the desired tolerance. The error is well-controlled and convergent.
	
	At a fixed error threshold, $r_{\rm cut} \sim \xi^{-1}$ and $k_{\rm cut} \sim \xi$. The operation count for the real space sum of the mobility calculation is $n_{R} \sim C_{R} \, n_{\rm neigh} \, N$. The average number of neighbors within a radius of $r_{\rm cut}$ of each particle is $n_{\rm neigh} \sim \phi \, r_{\rm cut}^{d_{f}}\sim \phi \, \xi^{-d_{f}}$ in a dispersion with volume fraction $\phi$ and fractal dimension $d_{f}$, so $n_{R} \sim C_{R} \, \phi \, \xi^{d_{f}} \, N$. $C_{R}$ is an implementation-specific constant that is independent of $N$, $\xi$, and $\phi$ to within hardware specific effects such as atomic conflicts, memory bandwidth saturation, etc. Drawing the real space random samples requires $n_{\rm iter}$ iterations to compute $\mathcal{M}_{R}^{1/2}\cdot d{\bf W}_{1}$ to within a given tolerance, so the operation count is $n_{R}^{\prime} \sim n_{\rm iter} \, n_{R} = n_{\rm iter} \, C_{R} \, \phi \, \xi^{-d_{f}} \, N$. Because the condition number of $\mathcal{M}_{r}$ is bounded for finite $\xi$, $n_{\rm iter}$ is independent of $N$. 
	
	The operation count to evaluate the wave space sum or draw wave space random samples is $n_{W} \sim C_{W} \, (\xi^{3} \, \phi \, N) \, \log (\xi^{3} \, \phi \, N) + C_{Q}\,NP^{3}$. The Fourier transform cost is controlled by $\xi\,\phi^{-1/3}\,N^{1/3}$, the number of Fourier modes per spatial dimension. The quantity $C_{Q}\,NP^{3}$ represents the operation count associated with the Spectral Ewald particle-to-grid and grid-to-particle quadrature operations. Like $C_{R}$, $C_{W}$ and $C_{Q}$ are implementation-specific and independent of $\xi$, $\phi$, and $N$ to within hardware effects. The total operation count is $n = n_{R}^{\prime} + n_{W}$, 
	\begin{align}
		n &\sim C_{R} \, n_{\rm iter} \, N \, \phi \, \xi^{-d_{f}} + C_{W} \, (\xi^{3} \, \phi^{-1} \, N) \, \log (\xi^{3} \, \phi^{-1} \, N) \nonumber \\
		&\quad + C_{Q}\,NP^{3}. \label{eqn:work}
	\end{align}
This expression makes it apparent that work can be freely partitioned between wave space and real space, with the total work optimized at a particular value of the splitting parameter, $\xi^{\ast}$. This optimal value is found by setting the derivative Equation \eqref{eqn:work} with respect to $\xi$ equal to zero. The solution is formally given by product logarithms, but can be simplified in the asymptotic limit of large $N$
	\begin{equation}
		\xi^{\ast} \sim \left(\frac{3C_{W}\,\log N }{C_{R}\,d_{f}\,n_{\rm iter}\,\phi^{2}}\right)^{\frac{-1}{d_{f}+3}}.
		\label{eqn:xistar}
	\end{equation}
	To leading order, the operation count associated with $\xi^{\ast}$ is
	\begin{align}
		n^{\ast} &\sim C_{R}\,n_{\rm iter} \left(\frac{3C_{W}}{d_{f}\,C_{R}\,n_{\rm iter}}\right)^{\frac{d_{f}}{d_{f}+3}} \, \phi^{\frac{3-d_{f}}{3+d_{f}}} \, N \, \left(\log N\right)^{\frac{d_{f}}{d_{f}+3}} \nonumber \\
		&\quad + C_{q} \, N P^{3}.
		\label{eqn:tstar}
	\end{align}
	For three-dimensional simulations, $ 1 \leq d_{f} \leq 3$, so the timing scales between $N\left(\log N\right)^{1/4}$ and $N\left(\log N\right)^{1/2}$. These estimates are nearly linear functions of the system size, and in practice the logarithmic factors are not observed. 

	\subsection{Comparison to the Force Coupling Method}
        
	The number of terms required to evaluate the wave space sum at a given error tolerance is dependent on the kernel used to represent the particle force distributions. The smallest length scale that needs to be resolved by the wave space sum is that of a single particle. Therefore, the wave space cutoff radius $k_{\rm cut}$ can be estimated by evaluating the error in the self mobility of a single particle relative to the stokes mobility ${\bf M}_{\rm stokes} = \left(6\pi\eta a\right)^{-1} {\bf I}$. Note that in practice a fixed grid spacing $~a/3$ is used in the FCM method,\cite{ForceCoupling_Lubrication} deemed to give sufficient accuracy. Given a shape factor for the particle in wave space, $s^{2}(k)$, the error in the approximate wave space mobility, $\widetilde{\bf M}_{self}^{(w)}$, is
	\begin{equation}
		\delta{\bf M} = {\bf M}_{self}^{(w)} - \widetilde{\bf M}_{self}^{(w)} = \frac{1}{\eta V} \sum_{k > k_{\rm cut}} s^{2}(k) \frac{1}{k^{2}}\left( {\bf I} - \hat{\bf k}\hat{\bf k} \right).
	\end{equation}
	The sum can be bounded by the integral
	\begin{align}
		\delta{\bf M} &\leq \frac{1}{\eta} \frac{1}{\left(2 \pi\right)^{3}} \int_{k > k_{\rm cut}} d{\bf k} \, s^{2}(k) \frac{1}{k^{2}}\left( {\bf I} - \hat{\bf k}\hat{\bf k} \right) \nonumber \\
		&= \frac{{\bf I}}{3\pi^{2}\eta}\int\limits_{k_{\rm cut}}^{\infty} dk \, s^{2}(k).
	\end{align}
	The relative error is given by
	\begin{equation}
		\epsilon_{k} = \frac{\left\lVert \delta{\bf M} \right\rVert}{\left\lVert {\bf M}_{\rm stokes} \right\rVert} \leq \frac{2a}{\pi}\int\limits_{k_{\rm cut}}^{\infty} dk \, s^{2}(k). \label{eqn:errbound}
	\end{equation}
	The positively split Ewald kernel is $s^{2}(k) = \left( \frac{\sin ka}{ka} \right)^{2} \left( 1 + \frac{k^{2}}{4\xi^{2}}\right)e^{-k^{2}/4\xi^{2}}$, and the FCM kernel is $s^{2}(k) = e^{-k^{2}a^{2}/\pi}$. The exponentials in the PSE and FCM kernels are equal when $\xi^{2}a^{2} = \pi/4$. 
        
\begin{figure}
	  \includegraphics{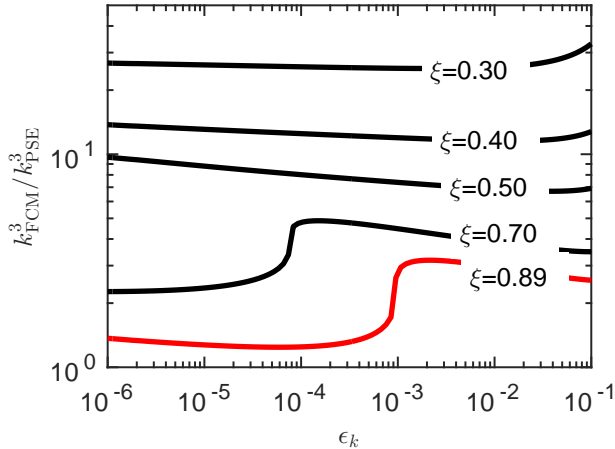}
	  \caption{Relative speedup for the wave space sum using the PSE kernel compared to the FCM kernel as a function of truncation error. The red line corresponds to $\xi^{2}=\pi/4a^{2}$, where the PSE and FCM exponentials have the same decay rate.}
	  \label{fig:FCMcompare}
	\end{figure}

        Direct comparison of computational performance between FCM and PSE is difficult because of the relative cost of the FFTs and the spreading and interpolations depends strongly on implementation specifics. For example, Yeo and Maxey\cite{ForceCoupling_Lubrication} report that in their implementation the FFT takes only 15\% of the total cost, which is otherwise dominated by spreading and interpolation, whereas we find that the FFT is the dominant cost in our implementation. Instead of trying to compare numerical performance directly, we compare here the size of the FFT grid required by the two methods in order to achieve a certain target tolerance. Specifically, we compare the minimum value of $k_{\rm cut}$ required to reach a set error tolerance for the two kernels. In three dimensions, the execution time of the FFT roughly scales as $k_{\rm cut}^{3}$, therefore $k_{\rm cut,FCM}^{3} / k_{\rm cut,PSE}^{3}$ provides an estimate of how much faster the FFT part of the algorithm will perform when using PSE. Figure \ref{fig:FCMcompare} shows this ratio as a function of error tolerance. For an error tolerance of $10^{-3}$, $k_{\rm cut,FCM}^{3} / k_{\rm cut,PSE}^{3}\approx 10$ for a typical value of ${\xi}a=0.5$ (see Fig. \ref{fig:timing}), i.e. the FFT grid in the PSE algorithm is ten times smaller than that in FCM for these parameters.
	
	This error estimate can also be used to provide a slightly tighter bound on the error than that presented by \citeauthor{lindbo-tornberg}\cite{lindbo-tornberg}. Although there is not an analytical form to Equation \eqref{eqn:errbound} for the PSE kernel, an analytical expression does exist for the approximation $s^{2}(k) \leq \frac{1}{\left(ka\right)^{2}} \left( 1 + \frac{k^{2}}{4\xi^{2}}\right)e^{-k^{2}/4\xi^{2}}$. That error estimate is
	\begin{equation}
		\epsilon_{k} \leq \frac{ 4k_{\rm cut}^{-1}e^{-k_{\rm cut}^{2}/4\xi^{2}} - \pi^{1/2}\xi^{-1}{\rm erfc}(k_{\rm cut}/2\xi) }{ 2\pi a }.
	\end{equation}
	
\section{Results and discussion}

	We have implemented our PSE method on GPUs using CUDA in order to take benefit of the massively parallel architecture. The real space Ewald summation is routinely accelerated with linked-cell lists and a specified truncation radius for the interactions. This approach can reduce the cost of the sum from $\mathcal{O}\left(N^{2}\right)$ to $\mathcal{O}\left(N\right)$. Our implementation of the real space sum uses state-of-the-art cell and neighbor list algorithms supplied in the HOOMD-blue software suite \cite{hoomd1,hoomd2}. The complicated real space mobility functions are pre-tabulated and stored in textured memory on the GPU, so their evaluation in the mobility multiplication is reduced to a memory fetch. The wave space sum is also truncated, and is accelerated using the non-uniform Fast Fourier transform (NUFFT).\cite{NUFFT}
\comment{The most common NUFFT in particle simulation is particle-mesh Ewald, but improved methods have recently been developed \cite{lindbo-tornberg}. NUFFT algorithms evaluate the wave space summation in $\mathcal{O}\left(N \log N\right)$ operations.}
Our NUFFT implementation is based on the Spectral Ewald technique of \citeauthor{lindbo-tornberg}\cite{lindbo-tornberg} and uses the CUDA library CUFFT to efficiently evaluate the FFTs. The spreading operator ${\bf S}$ uses atomic additions and a particle-based (i.e., one block of threads per particle) algorithm to transfer particle forces to the grid. A particle-based algorithm is also used for ${\bf S}^{\dagger}$ to interpolate velocities from the grid to the particles. The atomic operations are very fast so particle-based approaches to spreading outperform grid-based approaches.\cite{chow-spread}

The wave space parts of the deterministic and stochastic calculations are combined to minimize the number of FFTs required. Particle forces are spread to the real space grid, and transformed to wave space with one set of FFTs. After projecting the wave space gridded forces to the wave space velocities, the Brownian velocity is added directly to the grid. A set of inverse FFTs produces the real space velocity distribution on the grid, which is then interpolated onto particles. In principle, different values of $\xi$ can be used for the stochastic and deterministic calculations, in which case the FFTs could not be combined and an additional set of FFTs would be required to generate the Brownian displacements.

\begin{figure}
  \includegraphics{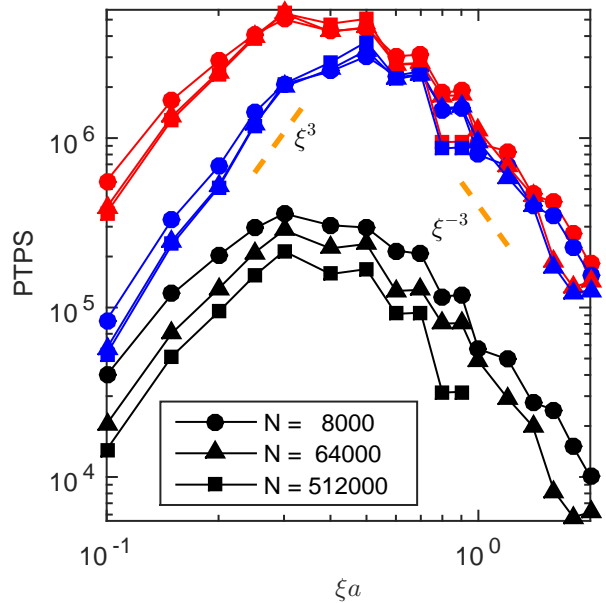}
  \caption{Timing results for calculations on a random suspension of hard spheres ($\phi=0.1$). The cost of the computation in terms of particle timesteps per second (PTPS) is given for computing $\mathcal{M}\cdot{\bf F}$ (red), Brownian displacements calculation $\mathcal{M}^{1/2} \cdot {\bf N}$ using the positively split Ewald sum (blue), and Brownian calculation using the Lanczos method applied to the full mobility (black), where a single matrix-vector product costs the same as the red curve.}
  \label{fig:timing}
\end{figure}

The performance of the PSE algorithm is characterized in Figure \ref{fig:timing} for a hard sphere dispersion with $\phi = 0.1$ as a function of $\xi$ and $N$. The quantity reported, particle timesteps per second (PTPS), is the number of particles divided by the average time required to perform one step of the Euler-Maryuama time stepping method using the PSE algorithm, and represents the throughput of the algorithm on a per particle basis. The algorithm has been implemented as a plug-in to the GPU-based molecular dynamics tool kit HOOMD-Blue \cite{hoomd1,hoomd2}. All timing data were obtained using an NVIDIA Tesla K40 GPU with the Maxwell architecture using a relative error tolerance of $10^{-3}$ for both the Ewald splitting and the Lanczos iteration for computing the real space contribution to the Brownian displacements, unless otherwise noted. Drawing a sample of the Brownian displacement using the positively split Ewald kernel (blue lines) is nearly as fast as a mobility calculation $\mathcal{M}\cdot{\bf F}$ (red lines) and more than an order of magnitude faster than samples drawn using the Lanczos iterative scheme applied to the full mobility tensor (black lines). The real space and wave space parts of the algorithm scale as expected, as shown by the constant slopes $\xi^{3}$ and $\xi^{-3}$ (orange lines).

\begin{figure}
  \includegraphics{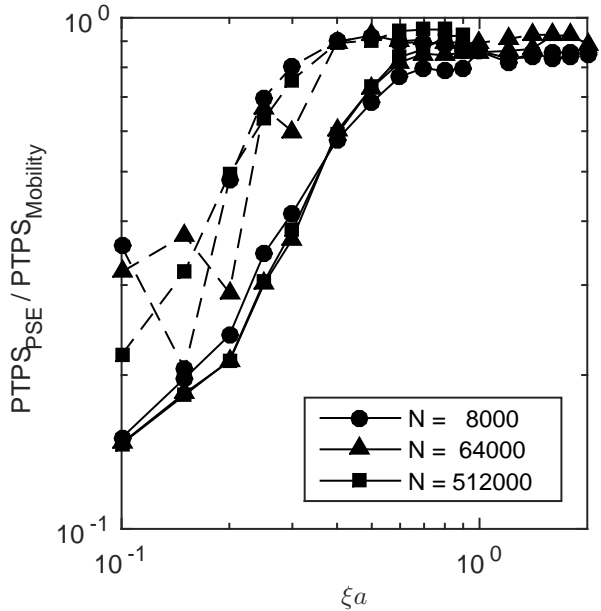}
  \caption{Speed of the stochastic calculation relative to a single mobility evaluation for a random suspension of hard spheres ($\phi=0.1$). Solid lines indicate calculations performed with a relative error tolerance of $10^{-3}$, dashed lines a relative error of $10^{-5}$.}
  \label{fig:RelMob}
\end{figure}

\begin{figure}
  \includegraphics{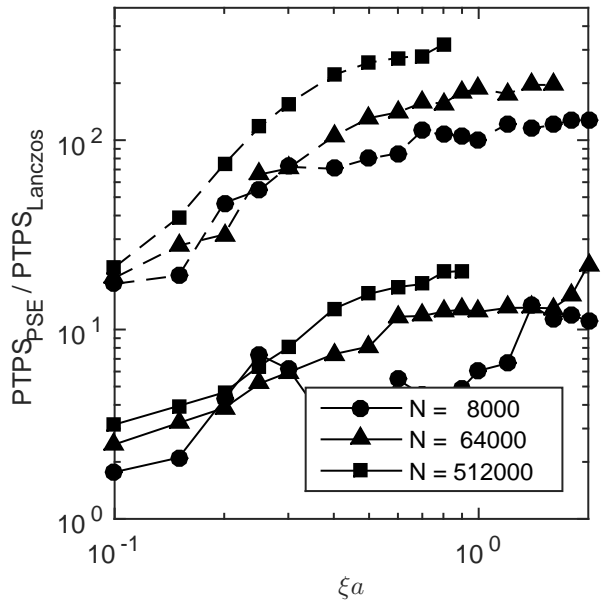}
  \caption{Speed of the PSE stochastic calculation relative to the iterative Lanczos stochastic calculation for a random suspension of hard spheres ($\phi=0.1$). Solid lines indicate calculations performed with a relative error tolerance of $10^{-3}$, dashed lines a relative error of $10^{-5}$.}
  \label{fig:RelLanc}
\end{figure}

Figure \ref{fig:timing} is an example of the algorithms' performance for a specific configuration. Systematically comparing the relative performance of the algorithms provides more insight into the advantage of the PSE method. The speed of sampling the random displacements $\mathcal{M}^{1/2} \cdot {\bf N}$, where ${\bf N}$ is a vector of i.i.d. standard Gaussian variables, using the PSE method relative to a single mobility product $\mathcal{M}\cdot{\bf F}$ is shown in Figure \ref{fig:RelMob}. The number of real space iterations required to sample the Brownian displacement increases weakly as $\xi$ decreases, but is independent of $N$. The PSE sampling speed relative to the Lanczos iteration sampling speed is shown in Figure \ref{fig:RelLanc}. The Lanczos approach applied to the full mobility becomes less efficient with increasing system size because the number of iterations required grows with $N$. Applying the Lanczos approach to draw samples from the the real space part of the mobility mitigates this problem because the real space operator has a condition number that is independent of $N$. The long range hydrodynamic interactions responsible for poor conditioning of the mobility are accounted for in the wave space part of the mobility. Samples are drawn from this part directly with no iterations required. In the stochastic PSE algorithm, the extra cost associated with the real space Lanczos iterations is compensated by increasing $\xi$ in order to shift work into the wave space part of the calculation. Figure \ref{fig:timing} demonstrates this as a difference in the location of the peak for the mobility product (red, ${\xi}a \approx 0.3$) and the random sampling (blue, ${\xi}a \approx 0.5$).

\begin{figure}
  \includegraphics{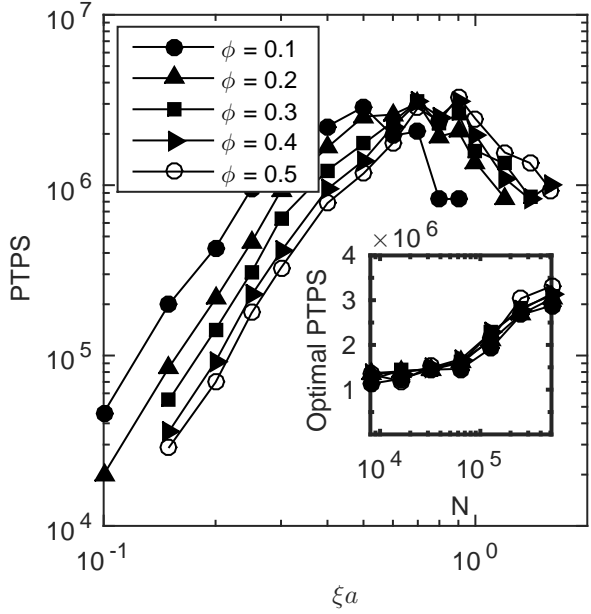}
  \caption{Performance of the PSE algorithm as a function of $\xi$ for a hard sphere dispersion of 512,000 particles at different volume fractions. The inset shows the optimal calculation performance as a function of $N$ for each volume fraction. }
  \label{fig:phiall}
\end{figure}

Figure \ref{fig:phiall} shows the performance of the PSE algorithm for 512,000 particles at different volume fractions. The location of the optima shifts to higher $\xi$ as the volume fraction increases, but the value at the optima remains unchanged. This is is further demonstrated in the inset, which shows the optimal performance as a function of system size for each $\phi$. It is clear that optimal calculation speed is independent of volume fraction at a given $N$. At higher $\phi$ each particle has more interactions at a given cutoff radius, so some work must be shifted to wave space to maintain the optimal speed. The increase in PTPS with $N$ is an artifact of the GPU. For smaller numbers of particles the hardware is not fully utilized. Increasing $N$ occupies more GPU cores and it hides latency times, resulting in the observed throughput increase. As shown in Table \ref{tab:tab1}, large enough systems exhibit the expected levelling off and mild, logarithmic decrease in simulation throughput once the number of particles is large enough to fully occupy the hardware.

\setlength{\tabcolsep}{3.25pt}
\begin{table}
	\caption{Performance results for calculations of  deterministic ($\mathcal{M}\cdot{\bf F}$) and stochastic ($\mathcal{M}^{1/2} \cdot {\bf N}$) velocities for a random suspension of hard spheres ($\phi=0.1$) at ${\xi}a=0.5$, which is roughly optimal for all $N$. $N$ is given in thousand of particles, values are given in terms of millions of particle timesteps per second ($10^{6}$ PTPS). }
	\begin{tabular}{lccccccccc}
	\hline\hline
	$N \cdot 10^{-3}$ & $8$ & $16$ & $32$ & $64$ & $128$ & $256$ & $512$ & $1024$ & $2048$ \\ \hline
	Deterministic & 5.2 & 5.5 & 6.1 & 5.6 & 5.9 & 6.2 & 6.2 & 5.3 & 4.8 \\
	Stochastic    & 0.9 & 1.0 & 1.4 & 1.6 & 2.2 & 2.8 & 3.3 & 3.2 & 3.0 \\
	\hline\hline
	\end{tabular}
	\label{tab:tab1}
\end{table}

\begin{figure}
  \includegraphics{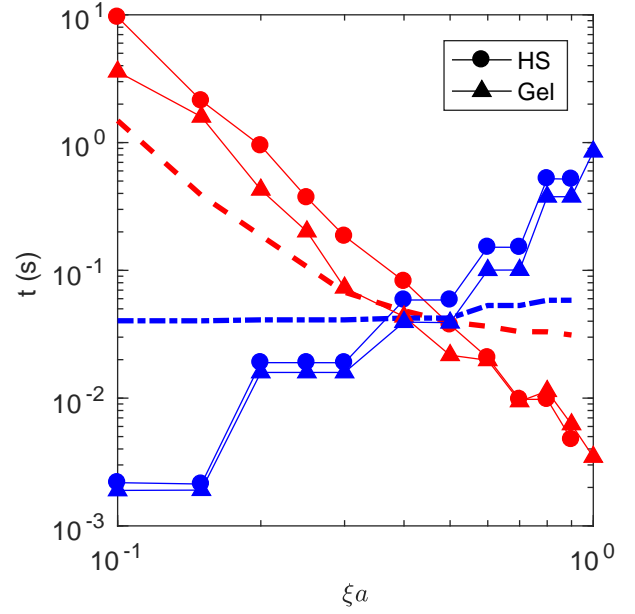}
  \caption{Execution time for real space (red) and wave space (blue) components of the stochastic calculation for hard sphere and gel suspensions of 512,000 particles at $\phi=0.1$. For the real space calculation, the solid curves are the time to compute the $n_{iter}$ real space multiplications, and the dashed line is the time required to perform all other calculations associated with the Lanczos iterations. For the wave space calculation, the solid line is the time to compute the FFT/iFFT pair, plus the wave space multiplication and the random number generation. The dashed-dotted line is the sum of the spreading and interpolation time. The total time to evaluate a real and wave space calculation is the sum of the dashed (dashed-dotted) and solid lines. }
  \label{fig:gel}
\end{figure}

Figure \ref{fig:gel} shows the time to draw the real space and wave space samples for a colloidal gel ($d_{f}\approx 2$) and a random hard sphere suspension ($d_{f}\approx 3$). Both systems have the same particle volume fraction, $\phi=0.10$, so they each fill the same amount of space, but the space is occupied differently because the gel has a lower fractal dimension than the hard sphere suspension. The real space sum becomes more efficient at lower fractal dimensions due to the decreased number of neighbors within a given radius, as demonstrated by the offset between the real space timing curves. The wave space calculation is largely unaffected by the fractal dimension, so the measured times for the hard sphere and gelled dispersion are nearly equal. In the present implementation, the peak algorithm performance is no different for different particle configurations. The effect of fractal dimension is seen in the real space calculation, but is masked near the optimum. Performance floors associated with auxiliary parts of the algorithm become relevant near the optimal speed. In wave space, the FFTs become nearly as fast as the to-grid and from-grid operations, so the peak performance is set by a balance of all three components. In real space, the evaluation of the near field sums is highly optimized by using the HOOMD linked-list implementation and by pre-tabulating the complicated truncated RPY kernels $F$ and $G$. The floor is set by the time required to evaluate inner products associated with the Lanczos iterates.

The algorithm is so efficient that further speedup of the wave space sum would need to be achieved by improving the to-grid and from-grid operations. Fast to-grid and from-grid projection is an important computational problem. In particular, these operations have received attention for use in reconstructing MRI images through combinations of source-based and grid-based approaches to associate the non-uniform data with its grid support \cite{spread1,spread2,spread3,spread4}. On GPUs, source-based algorithms reduce to a thread or thread block per particle using atomic operations to add to the grid point. Grid-based approaches, sometimes called gather algorithms, assign a thread or thread block per grid point or region and search for particles within that grid point's vicinity. The appropriate algorithm for a given application depends on the ratio of the source density (in our case, particle density) to the grid point density. In the extreme case of many source and few grid points, grid-based approach are optimal. In the opposite extreme of many grid points and few sources, source-based approaches are optimal. Both approaches and combinations thereof are reasonable choices for the intermediate case. We have found that in our application the fastest spreading comes from using a block of threads for each particle to assign values to grid points within its support. 

The to-grid and from-grid operations can also be accelerated using compactly supported kernels, as in immersed boundary methods \cite{peskin-ib,peskin-kernel}. These kernels are computationally efficient because a particle's support is localized to only 3, 4, or 6 grid points in each direction. However, compact kernels are not translationally invariant, which becomes more apparent as the support shrinks. More importantly, the accuracy with which the RPY tensor is computed cannot be controlled independently. In the FIB method only the far field RPY behavior is captured up to the quadrupole contributions if a kernel with a suitable positive second moment is used.\cite{fib}

Improvements to the spreading operations would lower the wave space floor, accelerating the wave space calculation improving overall efficiency. This highlights one of the most important features of the PSE algorithm: positive splitting of the real and wave space operators provides the ability to shift the optimal parameters based on changes in application or implementation. Improvements in performance of either the real space or wave space part of the algorithm can be translated into an overall speed-up by appropriately varying $\xi$. For example, the real space calculation could be more aggressively parallelized by using multiple GPUs and a smaller value of $\xi$. Using $M$ GPUs can reduce the calculation time roughly by a factor of $M$ with efficient domain decomposition algorithms \cite{hoomd2}, and distributing the inner products and additions required of the Lanczos iteration across multiple GPUs also translates to a factor of $M$ increase in speed. The PSE algorithm can be optimized for both software and hardware implementations, with the overall execution speed approaching that of the fastest component of the calculation. 

\section{Conclusion}

In summary, a new formulation for the RPY mobility tensor has been derived for spherical particles in periodic geometries. When computed by Ewald summation, both the real space and wave space components of this tensor are symmetric, positive definite. Using this fact, a method to rapidly calculate Brownian displacements has been developed that represents samples of the Brownian displacement as independent contributions drawn from the real space and wave space parts of the mobility. The real space displacement is calculated using an iterative scheme to approximate the square root of the real space mobility tensor, while the wave space component is calculated directly with a single matrix multiplication using FFTs. The new technique, Positively Split Ewald sampling (PSE), is two orders of magnitude faster than the standard Chebyshev polynomial approach of \citeauthor{fixman}. Furthermore its performance can be tuned based on suspension microstructure and is consistent across a broad range of volume fractions. PSE is flexible and its performance can be easily optimized for various hardware and software implementations. The algorithm's efficiency enables dynamic simulations of $\mathcal{O}\left(10^{6}\right)$ particles in reasonable wall times on modern GPU hardware. This technique should find many applications in the polymer and soft matter communities, particularly for large scale simulation of self-assembling materials. An implementation of this algorithm, written as a plug-in for the molecular dynamics suite HOOMD-blue \cite{hoomd1,hoomd2}, has been made freely available as part of the supplementary material. 

Generalization of the PSE method is possible. In the present work, only the force-velocity coupling among particles was modeled. However, an extension of the RPY tensor to describe higher order force moments such as torques and stresslets is possible\cite{RPY_Shear_Wall,RPY_Periodic_Shear}. These higher order terms would enter the formulation as a different set of shape factors, $ s(ka) $, describing how flows due to higher order force distributions on particle surfaces are propagated by the fluid. Likewise, polydispersity may be incorporated directly by making the shape factor size dependent.  Notably, equation \eqref{eqn:Mwave} cannot be used to efficiently perform the wave space calculations. Instead, the polydisperse shape factors must be incorporated into the spreading and interpolation functions much as is done in FCM\cite{ForceCoupling_Lubrication}. This approach is not taken in the present work with equal sized particles as a slightly larger support for the spreading and interpolation operations is required, which mildly reduces overall performance. A generalization from spherical particles to ellipsoids may also be possible through determination of an appropriate set of shape factors. Encouragingly, an equivalent RPY tensor for ellipsoids in an unbounded geometry is known and can be formally extended to describe the motion and interactions among colloidal particles with high aspect ratios\cite{claeys}. Finally, extension of the method to non-periodic geometries deserves serious consideration. One possibility for achieving this is through application of the General Geometry Ewald Method, which splits the hydrodynamic interactions into local and non-local parts by a screening procedure similar to Ewald summation\cite{GGEM}. In this approach, a Stokes-like equation subject to non-periodic boundary conditions is used to represent the non-local part of the interaction, while an analytical form of the HI is used for local part.  Ensuring positivity of the local and non-local operators while controlling the numerical error is key to making such an approach competitive with FIB, which can also be applied in non-periodic geometries. 

As the scale of dynamic simulation of colloidal materials grows, careful consideration must be given to how local errors, which are precisely bounded in the PSE method, propagate into global errors. For example, in most simulations of colloids, particles are subject to a conservative, pair-wise force. These conservative interactions should yield no net force on interacting particle pairs, but truncation errors in the wave space part of the PSE method applied to the mobility calculation will yield translation of the center of mass of the pair at a rate proportional to prescribed numerical tolerance. A local superposition of such errors, as might occur in a phase separating dispersion, could lead to a sizable net force acting on a condensed region of particles and lead to incorrect prediction of phase separation kinetics. The PSE method is spectrally accurate, however, and a smaller local error tolerance may be chosen to limit global erros over the scale of interest.  In particle-mesh Ewald simulations of $\mathcal{O}(10^3) $ colloidal particles, local tolerances on the order of $ 10^{-3} $ have been standard\cite{seasd}. Dynamic simulations of larger colloidal dispersions may necessitate tighter tolerances to limit the accumulation of local errors, which could lead to spurious particle fluxes. Alternatively, it may be possible to develop rescaled forms of the spreading and interpolation operators in the Spectral Ewald approximation to ensure that opposing forces sum to zero in aggregate when discretized. This is left for future work.

\section{Supplementary Material}
 See supplementary material for a GPU implementation of the PSE algorithm built as a plugin for HOOMD. The supplementary material also includes a sample script to perform a dynamic simulation using the plugin, as well as results for the osmotic pressure of a hard sphere dispersion. 

\acknowledgements
J. Swan and A. Fiore gratefully acknowledge funding from the MIT Energy Initiative Shell Seed Fund and NSF Career Award CBET-1554398. A. Donev and F. Balboa were supported in part by the National Science Foundation under award DMS-1418706, as well as the U.S. Department of Energy Office of Science, Office of Advanced Scientific Computing Research, Applied Mathematics program under Award Number DE-SC0008271.  Conversations with Edmond Chow are gratefully acknowledged.

\begin{appendix}

\section{Real Space Scalar Mobility Functions}\label{app:RealSpace}

	The scalar functions $F$ and $G$ are defined according to
	\begin{equation}
		{6\pi\eta{a}} \, {\bf M}^{(r)}_{ij} = F(r) \, \left( {\bf I} - \hat{\bf r}\hat{\bf r} \right) + G(r) \, \hat{\bf r}\hat{\bf r}
	\end{equation}
	where
	\begin{align}
		F(r) &= f_{0} + f_{1} \, e^{-r^2\xi^2} + f_{2} \, e^{-(r-2a)^2\xi^2} + f_{3} \, e^{-(r+2a)^2\xi^2} \nonumber \\
		&\quad + f_{4}  \, {\rm erfc}\left(r\xi\right) + f_{5} \, {\rm erfc}\left((r-2a)\xi\right) \nonumber \\
		&\quad + f_{6} \, {\rm erfc}\left((r+2a)\xi\right) \\
		G(r) &= g_{0} + g_{1} \, e^{-r^2\xi^2} + g_{2} \, e^{-(r-2a)^2\xi^2} + g_{3} \, e^{-(r+2a)^2\xi^2} \nonumber \\
		&\quad + g_{4}  \, {\rm erfc}\left(r\xi\right) + g_{5} \, {\rm erfc}\left((r-2a)\xi\right) \nonumber \\
		&\quad + g_{6} \, {\rm erfc}\left((r+2a)\xi\right)
	\end{align}
	\textbf{Case 1}, $r > 2a$
	
	First scalar mobility function:
	\begin{align*}
		f_{0} &= 0 \\
		f_{1} &= \frac{18 r^2 {\xi}^2+3}{64 \sqrt{\pi } a r^2 {\xi}^3} \\
		f_{2} &= \frac{2 {\xi}^2 (2 a-r) \left(4 a^2+4 a r+9 r^2\right)-2 a-3 r}{128 \sqrt{\pi } a r^3 {\xi}^3} \\
		f_{3} &= \frac{-2 {\xi}^2 (2 a+r) \left(4 a^2-4 a r+9 r^2\right)+2 a-3 r}{128 \sqrt{\pi } a r^3 {\xi}^3} \\
		f_{4} &= \frac{3-36 r^4 {\xi}^4}{128 a r^3 {\xi}^4} \\
		f_{5} &= \frac{4 {\xi}^4 (r-2 a)^2 \left(4 a^2+4 a r+9 r^2\right)-3}{256 a r^3 {\xi}^4} \\
		f_{6} &= \frac{4 {\xi}^4 (2 a+r)^2 \left(4 a^2-4 a r+9 r^2\right)-3}{256 a r^3 {\xi}^4}
	\end{align*}
	Second scalar mobility function:
	\begin{align*}
		g_{0} &= 0 \\
		g_{1} &= \frac{6 r^2 {\xi}^2-3}{32 \sqrt{\pi } a r^2 {\xi}^3} \\
		g_{2} &= \frac{-2 {\xi}^2 (r-2 a)^2 (2 a+3 r)+2 a+3 r}{64 \sqrt{\pi } a r^3 {\xi}^3} \\
		g_{3} &= \frac{2 {\xi}^2 (2 a+r)^2 (2 a-3 r)-2 a+3 r}{64 \sqrt{\pi } a r^3 {\xi}^3} \\
		g_{4} &= -\frac{3 \left(4 r^4 {\xi}^4+1\right)}{64 a r^3 {\xi}^4} \\
		g_{5} &= \frac{3-4 {\xi}^4 (2 a-r)^3 (2 a+3 r)}{128 a r^3 {\xi}^4} \\
		g_{6} &= \frac{3-4 {\xi}^4 (2 a-3 r) (2 a+r)^3}{128 a r^3 {\xi}^4}
	\end{align*}
	\textbf{Case 2}, $r \leq 2a$ 
	
	First scalar mobility function:
	\begin{align*}
		f_{0} &= -\frac{(r-2 a)^2 \left(4 a^2+4 a r+9 r^2\right)}{32 a r^3} \\
		f_{1} &= \frac{18 r^2 {\xi}^2+3}{64 \sqrt{\pi } a r^2 {\xi}^3} \\
		f_{2} &= \frac{2 {\xi}^2 (2 a-r) \left(4 a^2+4 a r+9 r^2\right)-2 a-3 r}{128 \sqrt{\pi } a r^3 {\xi}^3} \\
		f_{3} &= \frac{-2 {\xi}^2 (2 a+r) \left(4 a^2-4 a r+9 r^2\right)+2 a-3 r}{128 \sqrt{\pi } a r^3 {\xi}^3} \\
		f_{4} &= \frac{3-36 r^4 {\xi}^4}{128 a r^3 {\xi}^4} \\
		f_{5} &= \frac{4 {\xi}^4 (r-2 a)^2 \left(4 a^2+4 a r+9 r^2\right)-3}{256 a r^3 {\xi}^4} \\
		f_{6} &= \frac{4 {\xi}^4 (2 a+r)^2 \left(4 a^2-4 a r+9 r^2\right)-3}{256 a r^3 {\xi}^4} 
	\end{align*}
	Second scalar mobility function:
	\begin{align*}
		g_{0} &= \frac{(2 a-r)^3 (2 a+3 r)}{16 a r^3} \\
		g_{1} &= \frac{6 r^2 {\xi}^2-3}{32 \sqrt{\pi } a r^2 {\xi}^3} \\
		g_{2} &= \frac{-2 {\xi}^2 (r-2 a)^2 (2 a+3 r)+2 a+3 r}{64 \sqrt{\pi } a r^3 {\xi}^3} \\
		g_{3} &= \frac{2 {\xi}^2 (2 a+r)^2 (2 a-3 r)-2 a+3 r}{64 \sqrt{\pi } a r^3 {\xi}^3} \\
		g_{4} &= -\frac{3 \left(4 r^4 {\xi}^4+1\right)}{64 a r^3 {\xi}^4} \\
		g_{5} &= \frac{3-4 {\xi}^4 (2 a-r)^3 (2 a+3 r)}{128 a r^3 {\xi}^4} \\
		g_{6} &= \frac{3-4 {\xi}^4 (2 a-3 r) (2 a+r)^3}{128 a r^3 {\xi}^4}
	\end{align*}	
	\textbf{ Case 3, Self Mobility} The real space part of the self mobility of a particle is given by the limit of equation \eqref{eqn:Mreal} as $r\rightarrow 0$, specifically ${\bf M}^{(r)} = F(r{\rightarrow}0,\xi)\,{\bf I}$:
	\begin{equation}
		{6\pi{\eta}a}\,{\bf M}^{(r)}_{ii} = \frac{1}{4 \pi^{1/2} \xi a} \left( 1 - e^{-4a^{2}\xi^{2}} + 4 \pi^{1/2}a\xi \, {\rm erfc}\left(2a\xi\right)\right){\bf I}. \label{eqn:self}
	\end{equation}

\section{Proof of Positive-Definiteness}\label{app:SPD}

	Here we follow the proof presented by \citeauthor{cichocki_SPD}\cite{cichocki_SPD} for the positive-definiteness of a tensor defined by the quadratic form
	\begin{equation}
		\langle {\bf g} \, \lvert \, {\bf J} \, \rvert \, {\bf g} \rangle \equiv  \int d{\bf x} \, \int d{\bf y} \, {\bf g}^{\ast}\left({\bf x}\right) \cdot {\bf J}\left({\bf x},{\bf y}\right) \cdot {\bf g}\left({\bf y}\right)
	\end{equation}
	where an asterisk denotes the complex conjugate and the integrals in ${\bf x}$ and ${\bf y}$ are over all space, and \citeauthor{cichocki_SPD} define
	\begin{equation}
		{\bf g}\left( {\bf x} \right) \equiv \sum_{i} {\bf w}_{i}\left({\bf x}\right) \cdot {\bf d}_{i}
	\end{equation}
	where ${\bf w}_{i}$ is a tensor such that ${\bf w}_{i} \cdot {\bf F}_i$ is the force density on the surface of particle $i$ and ${\bf d}_{i}$ is an arbitrary vector.\cite{RPY_Periodic_Shear} In this work, ${\bf w}_{i}\left( {\bf x} \right) = \frac{1}{4\pi a^{2}} \, \delta\left( \lVert {\bf x} - {\bf x}_{i} \rVert - a \right) \, {\bf I}$, where ${\bf x}_{i}$ is the location of the center of particle $i$, and the pair mobility ${\bf M}_{ij}$ is
	\begin{equation}
		{\bf M}_{ij} = \langle {\bf w}_{i} \, \lvert \, {\bf J} \, \rvert \, {\bf w}_{j} \rangle.
	\end{equation}
	Because the Green's function ${\bf J}$ is positive-definite, and its quadratic form can be related to ${\bf M}_{ij}$, $\mathcal{M}$ can be shown to be positive-definite as well:
	\begin{equation}
		0 \leq \langle {\bf g} \, \lvert \, {\bf J} \, \rvert \, {\bf g} \rangle = \sum_{i,j} {\bf d}_{i}^{\ast} \cdot {\bf M}_{ij} \cdot {\bf d}_{j}.
	\end{equation}
	
	The wave space representation of ${\bf J}$ is ${\bf J} = \sum_{{\bf k}\neq{\bf 0}} {\bf J}_{\bf k}$, where $0 \leq \langle {\bf g} \, \lvert \, {\bf J}_{\bf k} \, \rvert \, {\bf g} \rangle$. In this representation, the mobility is
	\begin{equation}
		{\bf M}_{ij} = \langle {\bf w}_{i} \, \lvert \, \sum\limits_{{\bf k}\neq{\bf 0}}{\bf J}_{\bf k} \, \rvert \, {\bf w}_{j} \rangle = \sum\limits_{{\bf k}\neq{\bf 0}} \, \langle {\bf w}_{i} \, \lvert \, {\bf J}_{\bf k} \, \rvert \, {\bf w}_{j} \rangle,
	\end{equation}	
	which can be decomposed into separate sums with the splitting function (homotopy) $H(k)$ to yield
	\begin{equation}
		{\bf M}_{ij} = \sum_{{\bf k}\neq{\bf 0}} \, \langle {\bf w}_{i} \, \lvert \, \left(1-H(k)\right)\,{\bf J}_{\bf k} \, \rvert \, {\bf w}_{j} \rangle + \sum_{{\bf k}\neq{\bf 0}} \, \langle {\bf w}_{i} \, \lvert \,  H(k)\,{\bf J}_{\bf k} \, \rvert \, {\bf w}_{j} \rangle,
	\end{equation}
	where the first term is the real space component of the mobility and the second term is the wave space component. It follows from $0 \leq H \leq 1$ and $0 \leq \langle {\bf g} \, \lvert \, {\bf J}_{\bf k} \, \rvert \, {\bf g} \rangle$ that
	\begin{align}
		0 \leq \sum_{{\bf k}\neq{\bf 0}} \langle {\bf g} \, \lvert \, (1-H(k)) \, {\bf J}_{\bf k} \, \rvert \, {\bf g} \rangle =  \sum_{i,k} {\bf d}_{i}^{\ast} \cdot {\bf M}^{(r)}_{ij} \cdot {\bf d}_{j}\\
		0 \leq \sum_{{\bf k}\neq{\bf 0}} \langle {\bf g} \, \lvert \, H(k) \, {\bf J}_{\bf k} \, \rvert \, {\bf g} \rangle = \sum_{i,j} {\bf d}_{i}^{\ast} \cdot {\bf M}^{(w)}_{ij} \cdot {\bf d}_{j}
	\end{align}
	which completes the proof that ${\bf M}^{(r)}_{ij}$ and ${\bf M}^{(w)}_{ij}$ are independently positive-definite. Note that \textit{any} choice of $H$ such that $ 0 \leq H \leq 1 $ will produce a positive splitting, provided that each ${\bf J}_{\bf k}$ is positive definite. 
\end{appendix}

\begin{thebibliography}{44}%
\makeatletter
\providecommand \@ifxundefined [1]{%
 \@ifx{#1\undefined}
}%
\providecommand \@ifnum [1]{%
 \ifnum #1\expandafter \@firstoftwo
 \else \expandafter \@secondoftwo
 \fi
}%
\providecommand \@ifx [1]{%
 \ifx #1\expandafter \@firstoftwo
 \else \expandafter \@secondoftwo
 \fi
}%
\providecommand \natexlab [1]{#1}%
\providecommand \enquote  [1]{``#1''}%
\providecommand \bibnamefont  [1]{#1}%
\providecommand \bibfnamefont [1]{#1}%
\providecommand \citenamefont [1]{#1}%
\providecommand \href@noop [0]{\@secondoftwo}%
\providecommand \href [0]{\begingroup \@sanitize@url \@href}%
\providecommand \@href[1]{\@@startlink{#1}\@@href}%
\providecommand \@@href[1]{\endgroup#1\@@endlink}%
\providecommand \@sanitize@url [0]{\catcode `\\12\catcode `\$12\catcode
  `\&12\catcode `\#12\catcode `\^12\catcode `\_12\catcode `\%12\relax}%
\providecommand \@@startlink[1]{}%
\providecommand \@@endlink[0]{}%
\providecommand \url  [0]{\begingroup\@sanitize@url \@url }%
\providecommand \@url [1]{\endgroup\@href {#1}{\urlprefix }}%
\providecommand \urlprefix  [0]{URL }%
\providecommand \Eprint [0]{\href }%
\providecommand \doibase [0]{http://dx.doi.org/}%
\providecommand \selectlanguage [0]{\@gobble}%
\providecommand \bibinfo  [0]{\@secondoftwo}%
\providecommand \bibfield  [0]{\@secondoftwo}%
\providecommand \translation [1]{[#1]}%
\providecommand \BibitemOpen [0]{}%
\providecommand \bibitemStop [0]{}%
\providecommand \bibitemNoStop [0]{.\EOS\space}%
\providecommand \EOS [0]{\spacefactor3000\relax}%
\providecommand \BibitemShut  [1]{\csname bibitem#1\endcsname}%
\let\auto@bib@innerbib\@empty
\bibitem [{\citenamefont {Swan}\ and\ \citenamefont {Brady}(2011)}]{swan2011}%
  \BibitemOpen
  \bibfield  {author} {\bibinfo {author} {\bibfnamefont {J.~W.}\ \bibnamefont
  {Swan}}\ and\ \bibinfo {author} {\bibfnamefont {J.~F.}\ \bibnamefont
  {Brady}},\ }\href@noop {} {\bibfield  {journal} {\bibinfo  {journal} {The
  Journal of chemical physics}\ }\textbf {\bibinfo {volume} {135}},\ \bibinfo
  {pages} {014701} (\bibinfo {year} {2011})}\BibitemShut {NoStop}%
\bibitem [{\citenamefont {Varga}, \citenamefont {Wang},\ and\ \citenamefont
  {Swan}(2015)}]{zsigi}%
  \BibitemOpen
  \bibfield  {author} {\bibinfo {author} {\bibfnamefont {Z.}~\bibnamefont
  {Varga}}, \bibinfo {author} {\bibfnamefont {G.}~\bibnamefont {Wang}}, \ and\
  \bibinfo {author} {\bibfnamefont {J.}~\bibnamefont {Swan}},\ }\href {\doibase
  10.1039/C5SM01414J} {\bibfield  {journal} {\bibinfo  {journal} {Soft Matter}\
  }\textbf {\bibinfo {volume} {11}},\ \bibinfo {pages} {9009} (\bibinfo {year}
  {2015})}\BibitemShut {NoStop}%
\bibitem [{\citenamefont {Varga}\ and\ \citenamefont {Swan}(2016)}]{zsigi2}%
  \BibitemOpen
  \bibfield  {author} {\bibinfo {author} {\bibfnamefont {Z.}~\bibnamefont
  {Varga}}\ and\ \bibinfo {author} {\bibfnamefont {J.}~\bibnamefont {Swan}},\
  }\href {\doibase 10.1039/C6SM01285J} {\bibfield  {journal} {\bibinfo
  {journal} {Soft Matter}\ }\textbf {\bibinfo {volume} {12}},\ \bibinfo {pages}
  {7670} (\bibinfo {year} {2016})}\BibitemShut {NoStop}%
\bibitem [{\citenamefont {Ladd}(1996)}]{ladd_sedimentation}%
  \BibitemOpen
  \bibfield  {author} {\bibinfo {author} {\bibfnamefont {A.~J.}\ \bibnamefont
  {Ladd}},\ }\href@noop {} {\bibfield  {journal} {\bibinfo  {journal} {Physical
  review letters}\ }\textbf {\bibinfo {volume} {76}},\ \bibinfo {pages} {1392}
  (\bibinfo {year} {1996})}\BibitemShut {NoStop}%
\bibitem [{\citenamefont {Rotne}\ and\ \citenamefont {Prager}(1969)}]{rpy}%
  \BibitemOpen
  \bibfield  {author} {\bibinfo {author} {\bibfnamefont {J.}~\bibnamefont
  {Rotne}}\ and\ \bibinfo {author} {\bibfnamefont {S.}~\bibnamefont {Prager}},\
  }\href {\doibase http://dx.doi.org/10.1063/1.1670977} {\bibfield  {journal}
  {\bibinfo  {journal} {The Journal of Chemical Physics}\ }\textbf {\bibinfo
  {volume} {50}},\ \bibinfo {pages} {4831} (\bibinfo {year}
  {1969})}\BibitemShut {NoStop}%
\bibitem [{\citenamefont {Brady}\ and\ \citenamefont {Bossis}(1988)}]{sd}%
  \BibitemOpen
  \bibfield  {author} {\bibinfo {author} {\bibfnamefont {J.~F.}\ \bibnamefont
  {Brady}}\ and\ \bibinfo {author} {\bibfnamefont {G.}~\bibnamefont {Bossis}},\
  }\href@noop {} {\bibfield  {journal} {\bibinfo  {journal} {Annual review of
  fluid mechanics}\ }\textbf {\bibinfo {volume} {20}},\ \bibinfo {pages} {111}
  (\bibinfo {year} {1988})}\BibitemShut {NoStop}%
\bibitem [{\citenamefont {Liang}\ \emph {et~al.}(2013)\citenamefont {Liang},
  \citenamefont {Gimbutas}, \citenamefont {Greengard}, \citenamefont {Huang},\
  and\ \citenamefont {Jiang}}]{RPY_FMM}%
  \BibitemOpen
  \bibfield  {author} {\bibinfo {author} {\bibfnamefont {Z.}~\bibnamefont
  {Liang}}, \bibinfo {author} {\bibfnamefont {Z.}~\bibnamefont {Gimbutas}},
  \bibinfo {author} {\bibfnamefont {L.}~\bibnamefont {Greengard}}, \bibinfo
  {author} {\bibfnamefont {J.}~\bibnamefont {Huang}}, \ and\ \bibinfo {author}
  {\bibfnamefont {S.}~\bibnamefont {Jiang}},\ }\href@noop {} {\bibfield
  {journal} {\bibinfo  {journal} {Journal of Computational Physics}\ }\textbf
  {\bibinfo {volume} {234}},\ \bibinfo {pages} {133} (\bibinfo {year}
  {2013})}\BibitemShut {NoStop}%
\bibitem [{\citenamefont {Fixman}(1986)}]{fixman}%
  \BibitemOpen
  \bibfield  {author} {\bibinfo {author} {\bibfnamefont {M.}~\bibnamefont
  {Fixman}},\ }\href {\doibase 10.1021/ma00158a043} {\bibfield  {journal}
  {\bibinfo  {journal} {Macromolecules}\ }\textbf {\bibinfo {volume} {19}},\
  \bibinfo {pages} {1204} (\bibinfo {year} {1986})}\BibitemShut {NoStop}%
\bibitem [{\citenamefont {Chow}\ and\ \citenamefont {Saad}(2014)}]{chow-saad}%
  \BibitemOpen
  \bibfield  {author} {\bibinfo {author} {\bibfnamefont {E.}~\bibnamefont
  {Chow}}\ and\ \bibinfo {author} {\bibfnamefont {Y.}~\bibnamefont {Saad}},\
  }\href {\doibase 10.1137/130920587} {\bibfield  {journal} {\bibinfo
  {journal} {SIAM Journal on Scientific Computing}\ }\textbf {\bibinfo {volume}
  {36}},\ \bibinfo {pages} {A588} (\bibinfo {year} {2014})}\BibitemShut
  {NoStop}%
\bibitem [{\citenamefont {Ando}\ \emph {et~al.}(2012)\citenamefont {Ando},
  \citenamefont {Chow}, \citenamefont {Saad},\ and\ \citenamefont
  {Skolnick}}]{ando-krylovHI}%
  \BibitemOpen
  \bibfield  {author} {\bibinfo {author} {\bibfnamefont {T.}~\bibnamefont
  {Ando}}, \bibinfo {author} {\bibfnamefont {E.}~\bibnamefont {Chow}}, \bibinfo
  {author} {\bibfnamefont {Y.}~\bibnamefont {Saad}}, \ and\ \bibinfo {author}
  {\bibfnamefont {J.}~\bibnamefont {Skolnick}},\ }\href {\doibase
  http://dx.doi.org/10.1063/1.4742347} {\bibfield  {journal} {\bibinfo
  {journal} {The Journal of Chemical Physics}\ }\textbf {\bibinfo {volume}
  {137}},\ \bibinfo {eid} {064106} (\bibinfo {year} {2012}),\
  http://dx.doi.org/10.1063/1.4742347}\BibitemShut {NoStop}%
\bibitem [{\citenamefont {Geyer}\ and\ \citenamefont {Winter}(2009)}]{tea}%
  \BibitemOpen
  \bibfield  {author} {\bibinfo {author} {\bibfnamefont {T.}~\bibnamefont
  {Geyer}}\ and\ \bibinfo {author} {\bibfnamefont {U.}~\bibnamefont {Winter}},\
  }\href {\doibase http://dx.doi.org/10.1063/1.3089668} {\bibfield  {journal}
  {\bibinfo  {journal} {The Journal of Chemical Physics}\ }\textbf {\bibinfo
  {volume} {130}},\ \bibinfo {eid} {114905} (\bibinfo {year} {2009}),\
  http://dx.doi.org/10.1063/1.3089668}\BibitemShut {NoStop}%
\bibitem [{\citenamefont {Ladd}, \citenamefont {Colvin},\ and\ \citenamefont
  {Frenkel}(1988)}]{lb1}%
  \BibitemOpen
  \bibfield  {author} {\bibinfo {author} {\bibfnamefont {A.~J.}\ \bibnamefont
  {Ladd}}, \bibinfo {author} {\bibfnamefont {M.~E.}\ \bibnamefont {Colvin}}, \
  and\ \bibinfo {author} {\bibfnamefont {D.}~\bibnamefont {Frenkel}},\
  }\href@noop {} {\bibfield  {journal} {\bibinfo  {journal} {Physical review
  letters}\ }\textbf {\bibinfo {volume} {60}},\ \bibinfo {pages} {975}
  (\bibinfo {year} {1988})}\BibitemShut {NoStop}%
\bibitem [{\citenamefont {Ladd}(1993)}]{lb2}%
  \BibitemOpen
  \bibfield  {author} {\bibinfo {author} {\bibfnamefont {A.~J.}\ \bibnamefont
  {Ladd}},\ }\href@noop {} {\bibfield  {journal} {\bibinfo  {journal} {Physical
  Review Letters}\ }\textbf {\bibinfo {volume} {70}},\ \bibinfo {pages} {1339}
  (\bibinfo {year} {1993})}\BibitemShut {NoStop}%
\bibitem [{\citenamefont {Ladd}\ and\ \citenamefont {Verberg}(2001)}]{lb3}%
  \BibitemOpen
  \bibfield  {author} {\bibinfo {author} {\bibfnamefont {A.}~\bibnamefont
  {Ladd}}\ and\ \bibinfo {author} {\bibfnamefont {R.}~\bibnamefont {Verberg}},\
  }\href@noop {} {\bibfield  {journal} {\bibinfo  {journal} {Journal of
  Statistical Physics}\ }\textbf {\bibinfo {volume} {104}},\ \bibinfo {pages}
  {1191} (\bibinfo {year} {2001})}\BibitemShut {NoStop}%
\bibitem [{\citenamefont {Gompper}\ \emph {et~al.}(2009)\citenamefont
  {Gompper}, \citenamefont {Ihle}, \citenamefont {Kroll},\ and\ \citenamefont
  {Winkler}}]{mpc}%
  \BibitemOpen
  \bibfield  {author} {\bibinfo {author} {\bibfnamefont {G.}~\bibnamefont
  {Gompper}}, \bibinfo {author} {\bibfnamefont {T.}~\bibnamefont {Ihle}},
  \bibinfo {author} {\bibfnamefont {D.}~\bibnamefont {Kroll}}, \ and\ \bibinfo
  {author} {\bibfnamefont {R.}~\bibnamefont {Winkler}},\ }in\ \href@noop {}
  {\emph {\bibinfo {booktitle} {Advanced computer simulation approaches for
  soft matter sciences III}}}\ (\bibinfo  {publisher} {Springer},\ \bibinfo
  {year} {2009})\ pp.\ \bibinfo {pages} {1--87}\BibitemShut {NoStop}%
\bibitem [{\citenamefont {Hoogerbrugge}\ and\ \citenamefont
  {Koelman}(1992)}]{dpd1}%
  \BibitemOpen
  \bibfield  {author} {\bibinfo {author} {\bibfnamefont {P.}~\bibnamefont
  {Hoogerbrugge}}\ and\ \bibinfo {author} {\bibfnamefont {J.}~\bibnamefont
  {Koelman}},\ }\href@noop {} {\bibfield  {journal} {\bibinfo  {journal} {EPL
  (Europhysics Letters)}\ }\textbf {\bibinfo {volume} {19}},\ \bibinfo {pages}
  {155} (\bibinfo {year} {1992})}\BibitemShut {NoStop}%
\bibitem [{\citenamefont {Warren}(1998)}]{dpd2}%
  \BibitemOpen
  \bibfield  {author} {\bibinfo {author} {\bibfnamefont {P.~B.}\ \bibnamefont
  {Warren}},\ }\href@noop {} {\bibfield  {journal} {\bibinfo  {journal}
  {Current opinion in colloid \& interface science}\ }\textbf {\bibinfo
  {volume} {3}},\ \bibinfo {pages} {620} (\bibinfo {year} {1998})}\BibitemShut
  {NoStop}%
\bibitem [{\citenamefont {Padding}\ and\ \citenamefont
  {Louis}(2006)}]{srd_scaling}%
  \BibitemOpen
  \bibfield  {author} {\bibinfo {author} {\bibfnamefont {J.~T.}\ \bibnamefont
  {Padding}}\ and\ \bibinfo {author} {\bibfnamefont {A.~A.}\ \bibnamefont
  {Louis}},\ }\href {\doibase 10.1103/PhysRevE.74.031402} {\bibfield  {journal}
  {\bibinfo  {journal} {Phys. Rev. E}\ }\textbf {\bibinfo {volume} {74}},\
  \bibinfo {pages} {031402} (\bibinfo {year} {2006})}\BibitemShut {NoStop}%
\bibitem [{\citenamefont {Delong}\ \emph {et~al.}(2014)\citenamefont {Delong},
  \citenamefont {Usabiaga}, \citenamefont {Delgado-Buscalioni}, \citenamefont
  {Griffith},\ and\ \citenamefont {Donev}}]{fib}%
  \BibitemOpen
  \bibfield  {author} {\bibinfo {author} {\bibfnamefont {S.}~\bibnamefont
  {Delong}}, \bibinfo {author} {\bibfnamefont {F.~B.}\ \bibnamefont
  {Usabiaga}}, \bibinfo {author} {\bibfnamefont {R.}~\bibnamefont
  {Delgado-Buscalioni}}, \bibinfo {author} {\bibfnamefont {B.~E.}\ \bibnamefont
  {Griffith}}, \ and\ \bibinfo {author} {\bibfnamefont {A.}~\bibnamefont
  {Donev}},\ }\href {\doibase http://dx.doi.org/10.1063/1.4869866} {\bibfield
  {journal} {\bibinfo  {journal} {The Journal of Chemical Physics}\ }\textbf
  {\bibinfo {volume} {140}},\ \bibinfo {eid} {134110} (\bibinfo {year}
  {2014}),\ http://dx.doi.org/10.1063/1.4869866}\BibitemShut {NoStop}%
\bibitem [{\citenamefont {Keaveny}(2014)}]{fcm}%
  \BibitemOpen
  \bibfield  {author} {\bibinfo {author} {\bibfnamefont {E.~E.}\ \bibnamefont
  {Keaveny}},\ }\href {\doibase http://dx.doi.org/10.1016/j.jcp.2014.03.013}
  {\bibfield  {journal} {\bibinfo  {journal} {Journal of Computational
  Physics}\ }\textbf {\bibinfo {volume} {269}},\ \bibinfo {pages} {61 }
  (\bibinfo {year} {2014})}\BibitemShut {NoStop}%
\bibitem [{\citenamefont {Yeo}\ and\ \citenamefont
  {Maxey}(2010)}]{ForceCoupling_Lubrication}%
  \BibitemOpen
  \bibfield  {author} {\bibinfo {author} {\bibfnamefont {K.}~\bibnamefont
  {Yeo}}\ and\ \bibinfo {author} {\bibfnamefont {M.~R.}\ \bibnamefont
  {Maxey}},\ }\href@noop {} {\bibfield  {journal} {\bibinfo  {journal} {Journal
  of computational physics}\ }\textbf {\bibinfo {volume} {229}},\ \bibinfo
  {pages} {2401} (\bibinfo {year} {2010})}\BibitemShut {NoStop}%
\bibitem [{\citenamefont {Delmotte}\ and\ \citenamefont
  {Keaveny}(2015)}]{FluctuatingFCM_DC}%
  \BibitemOpen
  \bibfield  {author} {\bibinfo {author} {\bibfnamefont {B.}~\bibnamefont
  {Delmotte}}\ and\ \bibinfo {author} {\bibfnamefont {E.~E.}\ \bibnamefont
  {Keaveny}},\ }\href@noop {} {\bibfield  {journal} {\bibinfo  {journal} {The
  Journal of chemical physics}\ }\textbf {\bibinfo {volume} {143}},\ \bibinfo
  {pages} {244109} (\bibinfo {year} {2015})}\BibitemShut {NoStop}%
\bibitem [{\citenamefont {Peskin}(2002)}]{peskin-ib}%
  \BibitemOpen
  \bibfield  {author} {\bibinfo {author} {\bibfnamefont {C.~S.}\ \bibnamefont
  {Peskin}},\ }\href@noop {} {\bibfield  {journal} {\bibinfo  {journal} {Acta
  numerica}\ }\textbf {\bibinfo {volume} {11}},\ \bibinfo {pages} {479}
  (\bibinfo {year} {2002})}\BibitemShut {NoStop}%
\bibitem [{\citenamefont {Lindbo}\ and\ \citenamefont
  {Tornberg}(2010)}]{lindbo-tornberg}%
  \BibitemOpen
  \bibfield  {author} {\bibinfo {author} {\bibfnamefont {D.}~\bibnamefont
  {Lindbo}}\ and\ \bibinfo {author} {\bibfnamefont {A.-K.}\ \bibnamefont
  {Tornberg}},\ }\href {\doibase http://dx.doi.org/10.1016/j.jcp.2010.08.026}
  {\bibfield  {journal} {\bibinfo  {journal} {Journal of Computational
  Physics}\ }\textbf {\bibinfo {volume} {229}},\ \bibinfo {pages} {8994 }
  (\bibinfo {year} {2010})}\BibitemShut {NoStop}%
\bibitem [{\citenamefont {Hasimoto}(1959)}]{hasimoto}%
  \BibitemOpen
  \bibfield  {author} {\bibinfo {author} {\bibfnamefont {H.}~\bibnamefont
  {Hasimoto}},\ }\href {\doibase 10.1017/S0022112059000222} {\bibfield
  {journal} {\bibinfo  {journal} {Journal of Fluid Mechanics}\ }\textbf
  {\bibinfo {volume} {5}},\ \bibinfo {pages} {317} (\bibinfo {year}
  {1959})}\BibitemShut {NoStop}%
\bibitem [{\citenamefont {Wang}\ and\ \citenamefont
  {Brady}(2016{\natexlab{a}})}]{SD_SpectralEwald}%
  \BibitemOpen
  \bibfield  {author} {\bibinfo {author} {\bibfnamefont {M.}~\bibnamefont
  {Wang}}\ and\ \bibinfo {author} {\bibfnamefont {J.~F.}\ \bibnamefont
  {Brady}},\ }\href@noop {} {\bibfield  {journal} {\bibinfo  {journal} {Journal
  of Computational Physics}\ }\textbf {\bibinfo {volume} {306}},\ \bibinfo
  {pages} {443} (\bibinfo {year} {2016}{\natexlab{a}})}\BibitemShut {NoStop}%
\bibitem [{\citenamefont {Wajnryb}\ \emph {et~al.}(2013)\citenamefont
  {Wajnryb}, \citenamefont {Mizerski}, \citenamefont {Zuk},\ and\ \citenamefont
  {Szymczak}}]{RPY_Shear_Wall}%
  \BibitemOpen
  \bibfield  {author} {\bibinfo {author} {\bibfnamefont {E.}~\bibnamefont
  {Wajnryb}}, \bibinfo {author} {\bibfnamefont {K.~A.}\ \bibnamefont
  {Mizerski}}, \bibinfo {author} {\bibfnamefont {P.~J.}\ \bibnamefont {Zuk}}, \
  and\ \bibinfo {author} {\bibfnamefont {P.}~\bibnamefont {Szymczak}},\
  }\href@noop {} {\bibfield  {journal} {\bibinfo  {journal} {Journal of Fluid
  Mechanics}\ }\textbf {\bibinfo {volume} {731}},\ \bibinfo {pages} {R3}
  (\bibinfo {year} {2013})}\BibitemShut {NoStop}%
\bibitem [{\citenamefont {Mizerski}\ \emph {et~al.}(2014)\citenamefont
  {Mizerski}, \citenamefont {Wajnryb}, \citenamefont {Zuk},\ and\ \citenamefont
  {Szymczak}}]{RPY_Periodic_Shear}%
  \BibitemOpen
  \bibfield  {author} {\bibinfo {author} {\bibfnamefont {K.~A.}\ \bibnamefont
  {Mizerski}}, \bibinfo {author} {\bibfnamefont {E.}~\bibnamefont {Wajnryb}},
  \bibinfo {author} {\bibfnamefont {P.~J.}\ \bibnamefont {Zuk}}, \ and\
  \bibinfo {author} {\bibfnamefont {P.}~\bibnamefont {Szymczak}},\ }\href@noop
  {} {\bibfield  {journal} {\bibinfo  {journal} {The Journal of chemical
  physics}\ }\textbf {\bibinfo {volume} {140}},\ \bibinfo {pages} {184103}
  (\bibinfo {year} {2014})}\BibitemShut {NoStop}%
\bibitem [{\citenamefont {Greengard}\ and\ \citenamefont {Lee}(2004)}]{NUFFT}%
  \BibitemOpen
  \bibfield  {author} {\bibinfo {author} {\bibfnamefont {L.}~\bibnamefont
  {Greengard}}\ and\ \bibinfo {author} {\bibfnamefont {J.}~\bibnamefont
  {Lee}},\ }\href {\doibase 10.1137/S003614450343200X} {\bibfield  {journal}
  {\bibinfo  {journal} {SIAM Review}\ }\textbf {\bibinfo {volume} {46}},\
  \bibinfo {pages} {443} (\bibinfo {year} {2004})}\BibitemShut {NoStop}%
\bibitem [{\citenamefont {Beenakker}(1986)}]{beenakker}%
  \BibitemOpen
  \bibfield  {author} {\bibinfo {author} {\bibfnamefont {C.~W.~J.}\
  \bibnamefont {Beenakker}},\ }\href {\doibase
  http://dx.doi.org/10.1063/1.451199} {\bibfield  {journal} {\bibinfo
  {journal} {The Journal of Chemical Physics}\ }\textbf {\bibinfo {volume}
  {85}},\ \bibinfo {pages} {1581} (\bibinfo {year} {1986})}\BibitemShut
  {NoStop}%
\bibitem [{\citenamefont {Darden}, \citenamefont {York},\ and\ \citenamefont
  {Pedersen}(1993)}]{pme}%
  \BibitemOpen
  \bibfield  {author} {\bibinfo {author} {\bibfnamefont {T.}~\bibnamefont
  {Darden}}, \bibinfo {author} {\bibfnamefont {D.}~\bibnamefont {York}}, \ and\
  \bibinfo {author} {\bibfnamefont {L.}~\bibnamefont {Pedersen}},\ }\href
  {\doibase http://dx.doi.org/10.1063/1.464397} {\bibfield  {journal} {\bibinfo
   {journal} {The Journal of Chemical Physics}\ }\textbf {\bibinfo {volume}
  {98}},\ \bibinfo {pages} {10089} (\bibinfo {year} {1993})}\BibitemShut
  {NoStop}%
\bibitem [{\citenamefont {Donev}\ and\ \citenamefont
  {Vanden-Eijnden}(2014)}]{DDFT_Hydro}%
  \BibitemOpen
  \bibfield  {author} {\bibinfo {author} {\bibfnamefont {A.}~\bibnamefont
  {Donev}}\ and\ \bibinfo {author} {\bibfnamefont {E.}~\bibnamefont
  {Vanden-Eijnden}},\ }\href {\doibase http://dx.doi.org/10.1063/1.4883520}
  {\bibfield  {journal} {\bibinfo  {journal} {J. Chem. Phys.}\ }\textbf
  {\bibinfo {volume} {140}},\ \bibinfo {eid} {234115} (\bibinfo {year}
  {2014})}\BibitemShut {NoStop}%
\bibitem [{\citenamefont {Bao}, \citenamefont {Kaye},\ and\ \citenamefont
  {Peskin}(2016)}]{peskin-kernel}%
  \BibitemOpen
  \bibfield  {author} {\bibinfo {author} {\bibfnamefont {Y.}~\bibnamefont
  {Bao}}, \bibinfo {author} {\bibfnamefont {J.}~\bibnamefont {Kaye}}, \ and\
  \bibinfo {author} {\bibfnamefont {C.~S.}\ \bibnamefont {Peskin}},\ }\href
  {\doibase http://dx.doi.org/10.1016/j.jcp.2016.04.024} {\bibfield  {journal}
  {\bibinfo  {journal} {Journal of Computational Physics}\ }\textbf {\bibinfo
  {volume} {316}},\ \bibinfo {pages} {139 } (\bibinfo {year}
  {2016})}\BibitemShut {NoStop}%
\bibitem [{\citenamefont {Anderson}, \citenamefont {Lorenz},\ and\
  \citenamefont {Travesset}(2008)}]{hoomd1}%
  \BibitemOpen
  \bibfield  {author} {\bibinfo {author} {\bibfnamefont {J.~A.}\ \bibnamefont
  {Anderson}}, \bibinfo {author} {\bibfnamefont {C.~D.}\ \bibnamefont
  {Lorenz}}, \ and\ \bibinfo {author} {\bibfnamefont {A.}~\bibnamefont
  {Travesset}},\ }\href {\doibase http://dx.doi.org/10.1016/j.jcp.2008.01.047}
  {\bibfield  {journal} {\bibinfo  {journal} {Journal of Computational
  Physics}\ }\textbf {\bibinfo {volume} {227}},\ \bibinfo {pages} {5342 }
  (\bibinfo {year} {2008})}\BibitemShut {NoStop}%
\bibitem [{\citenamefont {Glaser}\ \emph {et~al.}(2015)\citenamefont {Glaser},
  \citenamefont {Nguyen}, \citenamefont {Anderson}, \citenamefont {Lui},
  \citenamefont {Spiga}, \citenamefont {Millan}, \citenamefont {Morse},\ and\
  \citenamefont {Glotzer}}]{hoomd2}%
  \BibitemOpen
  \bibfield  {author} {\bibinfo {author} {\bibfnamefont {J.}~\bibnamefont
  {Glaser}}, \bibinfo {author} {\bibfnamefont {T.~D.}\ \bibnamefont {Nguyen}},
  \bibinfo {author} {\bibfnamefont {J.~A.}\ \bibnamefont {Anderson}}, \bibinfo
  {author} {\bibfnamefont {P.}~\bibnamefont {Lui}}, \bibinfo {author}
  {\bibfnamefont {F.}~\bibnamefont {Spiga}}, \bibinfo {author} {\bibfnamefont
  {J.~A.}\ \bibnamefont {Millan}}, \bibinfo {author} {\bibfnamefont {D.~C.}\
  \bibnamefont {Morse}}, \ and\ \bibinfo {author} {\bibfnamefont {S.~C.}\
  \bibnamefont {Glotzer}},\ }\href {\doibase
  http://dx.doi.org/10.1016/j.cpc.2015.02.028} {\bibfield  {journal} {\bibinfo
  {journal} {Computer Physics Communications}\ }\textbf {\bibinfo {volume}
  {192}},\ \bibinfo {pages} {97 } (\bibinfo {year} {2015})}\BibitemShut
  {NoStop}%
\bibitem [{\citenamefont {Guo}\ \emph {et~al.}(2015)\citenamefont {Guo},
  \citenamefont {Liu}, \citenamefont {Xu}, \citenamefont {Du},\ and\
  \citenamefont {Chow}}]{chow-spread}%
  \BibitemOpen
  \bibfield  {author} {\bibinfo {author} {\bibfnamefont {X.}~\bibnamefont
  {Guo}}, \bibinfo {author} {\bibfnamefont {X.}~\bibnamefont {Liu}}, \bibinfo
  {author} {\bibfnamefont {P.}~\bibnamefont {Xu}}, \bibinfo {author}
  {\bibfnamefont {Z.}~\bibnamefont {Du}}, \ and\ \bibinfo {author}
  {\bibfnamefont {E.}~\bibnamefont {Chow}},\ }\href@noop {} {\bibfield
  {journal} {\bibinfo  {journal} {Procedia Computer Science}\ }\textbf
  {\bibinfo {volume} {51}},\ \bibinfo {pages} {120} (\bibinfo {year}
  {2015})}\BibitemShut {NoStop}%
\bibitem [{\citenamefont {Stantchev}, \citenamefont {Dorland},\ and\
  \citenamefont {Gumerov}(2008)}]{spread1}%
  \BibitemOpen
  \bibfield  {author} {\bibinfo {author} {\bibfnamefont {G.}~\bibnamefont
  {Stantchev}}, \bibinfo {author} {\bibfnamefont {W.}~\bibnamefont {Dorland}},
  \ and\ \bibinfo {author} {\bibfnamefont {N.}~\bibnamefont {Gumerov}},\ }\href
  {\doibase http://dx.doi.org/10.1016/j.jpdc.2008.05.009} {\bibfield  {journal}
  {\bibinfo  {journal} {Journal of Parallel and Distributed Computing}\
  }\textbf {\bibinfo {volume} {68}},\ \bibinfo {pages} {1339 } (\bibinfo {year}
  {2008})},\ \bibinfo {note} {general-Purpose Processing using Graphics
  Processing Units}\BibitemShut {NoStop}%
\bibitem [{\citenamefont {Gai}\ \emph {et~al.}(2013)\citenamefont {Gai},
  \citenamefont {Obeid}, \citenamefont {Holtrop}, \citenamefont {Wu},
  \citenamefont {Lam}, \citenamefont {Fu}, \citenamefont {Haldar},
  \citenamefont {mei W.~Hwu}, \citenamefont {Liang},\ and\ \citenamefont
  {Sutton}}]{spread2}%
  \BibitemOpen
  \bibfield  {author} {\bibinfo {author} {\bibfnamefont {J.}~\bibnamefont
  {Gai}}, \bibinfo {author} {\bibfnamefont {N.}~\bibnamefont {Obeid}}, \bibinfo
  {author} {\bibfnamefont {J.~L.}\ \bibnamefont {Holtrop}}, \bibinfo {author}
  {\bibfnamefont {X.-L.}\ \bibnamefont {Wu}}, \bibinfo {author} {\bibfnamefont
  {F.}~\bibnamefont {Lam}}, \bibinfo {author} {\bibfnamefont {M.}~\bibnamefont
  {Fu}}, \bibinfo {author} {\bibfnamefont {J.~P.}\ \bibnamefont {Haldar}},
  \bibinfo {author} {\bibfnamefont {W.}~\bibnamefont {mei W.~Hwu}}, \bibinfo
  {author} {\bibfnamefont {Z.-P.}\ \bibnamefont {Liang}}, \ and\ \bibinfo
  {author} {\bibfnamefont {B.~P.}\ \bibnamefont {Sutton}},\ }\href {\doibase
  http://dx.doi.org/10.1016/j.jpdc.2013.01.001} {\bibfield  {journal} {\bibinfo
   {journal} {Journal of Parallel and Distributed Computing}\ }\textbf
  {\bibinfo {volume} {73}},\ \bibinfo {pages} {686 } (\bibinfo {year}
  {2013})}\BibitemShut {NoStop}%
\bibitem [{\citenamefont {Koziel}\ \emph {et~al.}(2015)\citenamefont {Koziel},
  \citenamefont {Leifsson}, \citenamefont {Lees}, \citenamefont
  {V.Krzhizhanovskaya}, \citenamefont {Dongarra}, \citenamefont {Sloot},
  \citenamefont {Guo}, \citenamefont {Liu}, \citenamefont {Xu}, \citenamefont
  {Du},\ and\ \citenamefont {Chow}}]{spread3}%
  \BibitemOpen
  \bibfield  {author} {\bibinfo {author} {\bibfnamefont {S.}~\bibnamefont
  {Koziel}}, \bibinfo {author} {\bibfnamefont {L.}~\bibnamefont {Leifsson}},
  \bibinfo {author} {\bibfnamefont {M.}~\bibnamefont {Lees}}, \bibinfo {author}
  {\bibfnamefont {V.}~\bibnamefont {V.Krzhizhanovskaya}}, \bibinfo {author}
  {\bibfnamefont {J.}~\bibnamefont {Dongarra}}, \bibinfo {author}
  {\bibfnamefont {P.~M.}\ \bibnamefont {Sloot}}, \bibinfo {author}
  {\bibfnamefont {X.}~\bibnamefont {Guo}}, \bibinfo {author} {\bibfnamefont
  {X.}~\bibnamefont {Liu}}, \bibinfo {author} {\bibfnamefont {P.}~\bibnamefont
  {Xu}}, \bibinfo {author} {\bibfnamefont {Z.}~\bibnamefont {Du}}, \ and\
  \bibinfo {author} {\bibfnamefont {E.}~\bibnamefont {Chow}},\ }\href {\doibase
  http://dx.doi.org/10.1016/j.procs.2015.05.210} {\bibfield  {journal}
  {\bibinfo  {journal} {Procedia Computer Science}\ }\textbf {\bibinfo {volume}
  {51}},\ \bibinfo {pages} {120 } (\bibinfo {year} {2015})}\BibitemShut
  {NoStop}%
\bibitem [{\citenamefont {S{\o}rensen}\ \emph {et~al.}(2008)\citenamefont
  {S{\o}rensen}, \citenamefont {Schaeffter}, \citenamefont {Noe},\ and\
  \citenamefont {Hansen}}]{spread4}%
  \BibitemOpen
  \bibfield  {author} {\bibinfo {author} {\bibfnamefont {T.~S.}\ \bibnamefont
  {S{\o}rensen}}, \bibinfo {author} {\bibfnamefont {T.}~\bibnamefont
  {Schaeffter}}, \bibinfo {author} {\bibfnamefont {K.~{\O}.}\ \bibnamefont
  {Noe}}, \ and\ \bibinfo {author} {\bibfnamefont {M.~S.}\ \bibnamefont
  {Hansen}},\ }\href@noop {} {\bibfield  {journal} {\bibinfo  {journal} {IEEE
  Transactions on Medical Imaging}\ }\textbf {\bibinfo {volume} {27}},\
  \bibinfo {pages} {538} (\bibinfo {year} {2008})}\BibitemShut {NoStop}%
\bibitem [{\citenamefont {Claeys}\ and\ \citenamefont {Brady}(1993)}]{claeys}%
  \BibitemOpen
  \bibfield  {author} {\bibinfo {author} {\bibfnamefont {I.~L.}\ \bibnamefont
  {Claeys}}\ and\ \bibinfo {author} {\bibfnamefont {J.~F.}\ \bibnamefont
  {Brady}},\ }\href@noop {} {\bibfield  {journal} {\bibinfo  {journal} {Journal
  of Fluid Mechanics}\ }\textbf {\bibinfo {volume} {251}},\ \bibinfo {pages}
  {411} (\bibinfo {year} {1993})}\BibitemShut {NoStop}%
\bibitem [{\citenamefont {Hern\'andez-Ortiz}, \citenamefont {de~Pablo},\ and\
  \citenamefont {Graham}(2007)}]{GGEM}%
  \BibitemOpen
  \bibfield  {author} {\bibinfo {author} {\bibfnamefont {J.~P.}\ \bibnamefont
  {Hern\'andez-Ortiz}}, \bibinfo {author} {\bibfnamefont {J.~J.}\ \bibnamefont
  {de~Pablo}}, \ and\ \bibinfo {author} {\bibfnamefont {M.~D.}\ \bibnamefont
  {Graham}},\ }\href {\doibase 10.1103/PhysRevLett.98.140602} {\bibfield
  {journal} {\bibinfo  {journal} {Phys. Rev. Lett.}\ }\textbf {\bibinfo
  {volume} {98}},\ \bibinfo {pages} {140602} (\bibinfo {year}
  {2007})}\BibitemShut {NoStop}%
\bibitem [{\citenamefont {Wang}\ and\ \citenamefont
  {Brady}(2016{\natexlab{b}})}]{seasd}%
  \BibitemOpen
  \bibfield  {author} {\bibinfo {author} {\bibfnamefont {M.}~\bibnamefont
  {Wang}}\ and\ \bibinfo {author} {\bibfnamefont {J.~F.}\ \bibnamefont
  {Brady}},\ }\href@noop {} {\bibfield  {journal} {\bibinfo  {journal} {Journal
  of Computational Physics}\ }\textbf {\bibinfo {volume} {306}},\ \bibinfo
  {pages} {443} (\bibinfo {year} {2016}{\natexlab{b}})}\BibitemShut {NoStop}%
\bibitem [{\citenamefont {Cichocki}\ \emph {et~al.}(2000)\citenamefont
  {Cichocki}, \citenamefont {Jones}, \citenamefont {Kutteh},\ and\
  \citenamefont {Wajnryb}}]{cichocki_SPD}%
  \BibitemOpen
  \bibfield  {author} {\bibinfo {author} {\bibfnamefont {B.}~\bibnamefont
  {Cichocki}}, \bibinfo {author} {\bibfnamefont {R.}~\bibnamefont {Jones}},
  \bibinfo {author} {\bibfnamefont {R.}~\bibnamefont {Kutteh}}, \ and\ \bibinfo
  {author} {\bibfnamefont {E.}~\bibnamefont {Wajnryb}},\ }\href@noop {}
  {\bibfield  {journal} {\bibinfo  {journal} {The Journal of Chemical Physics}\
  }\textbf {\bibinfo {volume} {112}},\ \bibinfo {pages} {2548} (\bibinfo {year}
  {2000})}\BibitemShut {NoStop}%
\end{thebibliography}
%

\end{document}